\documentclass[twocolumn]{aastex7}


\shorttitle{Dust-model Interpretation at kpc Scales}
\shortauthors{Guo et al.}
\submitjournal{ApJ}
\graphicspath{{./}{figures/}}

\begin{document}

\title{Mapping Dust Attenuation at Kiloparsec Scales. IV. A Dust-model Interpretation of Attenuation Curves in Nearby Galaxies}

\correspondingauthor{Cheng Li}
\email{cli2015@tsinghua.edu.cn}

\author[0009-0008-4962-665X]{Ruonan Guo}
\affiliation{Department of Astronomy, Tsinghua University, Beijing 100084, China}
\email{grn19@mails.tsinghua.edu.cn}

\author[0000-0002-8711-8970]{Cheng Li}
\affiliation{Department of Astronomy, Tsinghua University, Beijing 100084, China}
\email{cli2015@tsinghua.edu.cn}

\author[0009-0004-6271-4321]{Tao Jing}
\affiliation{Department of Astronomy, Tsinghua University, Beijing 100084, China}
\email{jingt20@mails.tsinghua.edu.cn}

\author[0000-0002-8999-6814]{Shuang Zhou}
\affiliation{INAF--Osservatorio Astronomico di Brera, via Brera, 28, 20159 Milano, Italy}
\email{shuang.zhou@inaf.it}

\author[0000-0002-0656-075X]{Niu Li}
\affiliation{National Astronomical Observatories, Chinese Academy of Sciences, Beijing 100101, China}
\email{liniu126@126.com}
\affiliation{Department of Astronomy, Tsinghua University, Beijing 100084, China}

\author[0009-0001-5437-410X]{Zhuo Cheng}
\affiliation{Department of Physics, The Chinese University of Hong Kong, Sha Tin, NT, Hong Kong, China}
\email{zhuocheng@cuhk.edu.hk}
\affiliation{Department of Astronomy, Tsinghua University, Beijing 100084, China}

\begin{abstract}
In this fourth paper on kiloparsec-scale dust attenuation, we ask whether the empirical trends found in Papers I--III can be translated into effective dust properties. Using attenuation curves for 2487 high-continuum-S/N spaxels in 91 SwiM v4.2 galaxies, we construct a grid of uniform-screen dust models composed of astronomical silicate and graphite grains with MRN-like size distributions. We fit the normalized attenuation-curve shape to constrain model parameters and then use the attenuation amplitude to estimate the model-dependent dust mass surface density. The inferred dust masses, compositions, and small-grain fractions are therefore effective quantities defined within the adopted attenuation model. The fitted models reproduce the main attenuation-curve variations and provide a direct bridge to Papers I--III: within this model, the relative 2175\AA\ bump sequence maps mainly onto the effective fraction of small graphitic/carbonaceous grains, while the NUV-slope sequence maps onto the effective small-silicate grain fraction and total silicate mass fraction. Non-SF regions have higher dust mass surface densities but lower dust-to-stellar mass ratios than SF regions, separating absolute dust content from dust content per unit stellar mass. Regions with larger specific H$\alpha$ surface brightness have larger dust-to-stellar mass ratios but lower inferred small-grain fractions, especially lower small-silicate fractions. In non-SF regions this quantity is interpreted as ionized-gas emission per unit stellar mass rather than as a direct sSFR. These model-dependent trends support a picture in which local dust processing changes the relative abundance of small grains and thereby shapes the attenuation-curve variations found across the series.
\end{abstract}



\section{Introduction} 

Interstellar dust is a small but influential component of the baryon cycle in galaxies. Although it accounts for less than 1\% of the interstellar medium (ISM) by mass, dust absorbs and scatters ultraviolet and optical photons, re-emits energy in the infrared and submillimeter, and participates in the chemical cycling of heavy elements \citep[e.g.][]{2018ARA&A..56..673G, 2020ARA&A..58..529S}. Stellar winds and supernovae inject metals and dust into the ISM; grains then grow, shatter, coagulate, and are destroyed as gas cycles between diffuse, dense, and star-forming phases. Dust therefore records both the enrichment history of galaxies and the local physical conditions of the ISM.

The wavelength dependence of dust extinction and attenuation provides one of the most direct observational links to dust properties. The total attenuation amplitude is sensitive to the dust column, while the curve shape depends on grain composition, grain-size distribution, scattering, and the relative geometry of stars and dust. Small grains are especially important in the ultraviolet: small carbonaceous grains are commonly associated with the 2175\AA\ feature \citep[e.g.][]{1993ApJ...414..632D,2023ApJ...948...55H}, whereas changes in the abundance of small grains can steepen or flatten the NUV curve. At the same time, attenuation curves are not pure extinction curves; they are effective quantities that mix intrinsic grain physics with radiative-transfer and stellar-population effects \citep[see the review by][]{2020ARA&A..58..529S}. Interpreting attenuation curves therefore requires both empirical measurements and controlled, explicitly stated model assumptions.

Most detailed constraints on dust composition and grain-size distributions come from the Milky Way, but resolved observations of nearby galaxies now make it possible to test dust physics across a wider range of environments. Broad-band and spectroscopic SED studies have measured dust masses and their correlations with stellar mass, star formation rate, gas mass, and metallicity in integrated galaxies and in sub-galactic regions \citep[e.g.][]{2010MNRAS.403.1894D, 2015MNRAS.452..397C, 2017MNRAS.464.4680D, 2014ApJ...780..172D, 2014ApJ...797...85G, 2017A&A...601A..55C, 2022A&A...668A.130C}. These studies usually rely on infrared dust emission and are powerful for estimating dust mass and temperature. A complementary approach is to use attenuation curves directly: if the attenuation curve is measured over enough wavelength range, its shape and amplitude can be compared with physical dust models to infer model-dependent dust mass, composition, and grain-size information.

This paper is the fourth in a series studying dust attenuation on kiloparsec scales by combining MaNGA optical integral-field spectroscopy, Swift/UVOT NUV imaging, and 2MASS near-infrared imaging. The first paper in the series \citep[hereafter Paper I]{2023ApJ...957...75Z} developed the attenuation-curve method and showed that both curve slopes and the 2175\AA\ bump vary substantially within nearby galaxies. The second paper \citep[hereafter Paper II]{2025arXiv250314028G} applied the same method to the larger SwiM v4.2 catalog and found that the optical opacity and NUV slope are governed mainly by local stellar-population and emission-line properties rather than by host-galaxy inclination or galactocentric radius. The third paper \citep[hereafter Paper III]{2026arXiv260710573G} focused on the 2175\AA\ bump and showed that its relative prominence is strongest at low specific H$\alpha$ surface brightness, $\Sigma_{H\alpha}/\Sigma_\ast$, especially in non-SF regions where this quantity traces ionized-gas emission per unit stellar mass rather than sSFR. Paper III also showed that the absolute bump amplitude partly follows dust column or surface density, whereas the normalized bump strengths better isolate the effective prominence of the bump carriers. Together, Papers I--III establish an empirical sequence in which local conditions shape the attenuation-curve amplitude, NUV slope, and bump strength. The goal of the present paper is to ask what effective dust properties can reproduce that sequence.

Here we use the attenuation curves measured in Paper II as input to a simple physical dust model. We assume a uniform dust-screen geometry and construct a library of silicate+graphite models with MRN-like grain-size distributions. We first compare normalized model curves with normalized observed attenuation curves to constrain the curve shape, and then use the absolute attenuation amplitude to estimate the dust mass surface density. From the best-matched models we derive several effective, model-dependent dust quantities: the dust mass surface density, the dust-to-stellar mass ratio, the silicate mass fraction, and the small-grain mass fractions in the silicate and graphite components. These quantities are used as physically motivated descriptors of the measured attenuation curves, not as a unique inversion of dust geometry, composition, and grain-size distribution. We then ask how these inferred quantities connect to the attenuation properties measured in Papers II and III and how they vary with local star-formation and ionized-gas diagnostics, stellar population age, metallicity, and SF/non-SF classification.

Section \ref{sec:data_method} summarizes the data, attenuation inputs, dust-model library, fitting procedure, and derived quantities. Section \ref{sec:results} presents the correlations between inferred dust properties and local galaxy properties, Section \ref{sec:discussion} discusses the implications, and Section \ref{sec:summary} summarizes the conclusions. We assume $\Omega_{M}=0.3$, $\Omega_{\Lambda}=0.7$, and $H_{0}=70\,\mathrm{km\,s^{-1}\,Mpc^{-1}}$.

\section{Data and Methodology}\label{sec:data_method}

\subsection{Data}\label{subsec:data}

We use the same MaNGA, Swift/UVOT, and 2MASS data products as Papers II and III. The parent catalog is SwiM\_v4.2 \citep{2023ApJS..268...63M}, which provides Swift/UVOT NUV imaging and MaNGA spectral maps matched to the UVOT {\tt uvw2} grid, with $1^{\prime\prime}$ pixels and a PSF FWHM of $2.92^{\prime\prime}$. For the present dust-model analysis, the relevant inputs are the MaNGA-based optical attenuation curve, the three UVOT attenuations in {\tt uvw2}, {\tt uvm2}, and {\tt uvw1}, and the 2MASS $K_s$ normalization used in Papers I and II. We refer readers to Papers II and III for the full survey description, photometric products, PSF matching, and 2MASS processing.

\subsection{\texorpdfstring{Input attenuation curves and local properties}{Input attenuation curves and local properties}}

The attenuation curves were obtained in Papers II and III by applying the Paper I method to the SwiM\_v4.2 sample. In brief, the method combines the optical relative attenuation curve measured from the MaNGA spectrum \citep{2020ApJ...896...38L}, an intrinsic NUV-to-NIR model spectrum from BIGS spectral fitting \citep{2019MNRAS.485.5256Z}, and the $K_s$-band normalization under the Paper I assumption that NIR attenuation is negligible. We use the resulting optical attenuation curve and UVOT attenuations, $A_{w2}$, $A_{m2}$, and $A_{w1}$, as constraints on the dust-model shape and amplitude. The larger NUV-only bump sample introduced in Paper III is not used, because the present model requires the full optical-to-NUV attenuation-curve shape and the $K_s$ normalization.

We also adopt the local stellar-population and emission-line quantities measured in Paper II, including $\Sigma_\ast$, $D_n4000$, EW(H$\delta_{\rm A}$), stellar ages and metallicities, $\Sigma_{H\alpha}$, sSFR, $\Sigma_{H\alpha}/\Sigma_\ast$, EW(H$\alpha$), gas-phase metallicity, and N2S2. Following Paper III, $\Sigma_{H\alpha}/\Sigma_\ast$ is interpreted as sSFR-like only in SF regions; in non-SF regions it is treated as ionized-gas emission per unit stellar mass.

For the dust-model sample we begin from the same attenuation-curve quality cuts used in Papers I--III, ${\rm SNR}_{\rm cont}>20$ and $A_V>0.25$. We additionally require non-negative UV attenuations in all three UVOT bands, acceptable continuum-fit quality, physically ordered optical-to-NUV curve shapes, and ${\rm SNR}>3$ in {\tt uvw2}, {\tt uvm2}, and {\tt uvw1}. These cuts reduce the 50,985 SwiM\_v4.2 spaxels in 551 galaxies to 2487 spaxels in 91 galaxies, including 1468 SF and 1019 non-SF spaxels. Spaxels are classified as star-forming (SF) or non-SF using the P1--P2 diagnostic diagram of \citet{2020MNRAS.499.5749J}, following Paper II. These cuts restrict the analysis to kpc-scale regions with robustly measurable attenuation curves, rather than to all spaxels in nearby galaxies.

\subsection{\texorpdfstring{Constructing the dust-model library}{Constructing the dust-model library}}

A dust attenuation model depends on intrinsic grain properties and the effective geometry between stars and dust. Because the geometry cannot be uniquely constrained for each kiloparsec-scale spaxel, we adopt a uniform foreground screen, consistent with the spectral-fitting assumption that all stellar populations experience the same attenuation. The dust mixture contains astronomical silicate and graphite grains, with material densities $\rho_{\rm sil}=3.5\,{\rm g\,cm^{-3}}$ and $\rho_{\rm gra}=2.24\,{\rm g\,cm^{-3}}$ and dielectric functions from \citet{1984ApJ...285...89D}.

For the grain-size distribution, we adopt the MRN power-law form, $dn/da\propto a^{-\alpha}$, over $[a_{\min},a_{\max}]$ \citep{1977ApJ...217..425M}. This simple form is appropriate because the constraints consist of one optical attenuation curve plus three UVOT attenuation points. In a uniform-screen model,
\begin{equation}
A_{\lambda}=1.086\tau_{\lambda}=1.086\times[\kappa_{abs}(\lambda)+\kappa_{sca}(\lambda)]\times\Sigma_{dust},
\end{equation}
where $\kappa_{\rm abs}$ and $\kappa_{\rm sca}$ are the mass absorption and scattering coefficients and $\Sigma_{\rm dust}$ is the dust mass surface density. We compute the coefficients with the BHMIE implementation of Mie theory \citep{1908AnP...330..377M,1983bhwd.book.....B}; for graphite, we use the ``$1/3$--$2/3$ approximation'' for the two dielectric responses relative to the basal plane \citep{1993ApJ...414..632D}.

A fully flexible two-component MRN model would have six size-distribution parameters plus one mixture parameter, which is too many for the available attenuation constraints. We therefore use four free parameters: a common maximum radius $a_{\rm sil,max}=a_{\rm gra,max}=a_{\max}$, two size-distribution slopes $\alpha_{\rm sil}$ and $\alpha_{\rm gra}$, and the silicate mass fraction $q_{\rm sil}$; the graphite fraction is $1-q_{\rm sil}$. The minimum radius is fixed to $a_{\min}=0.005\,\mu{\rm m}$. We precompute a grid with 33 values of $q_{\rm sil}$ over $[0.0,1.0]$, 29 values of $a_{\max}$ over $[0.025,2.0]\,\mu{\rm m}$, and 31 values each of $\alpha_{\rm sil}$ and $\alpha_{\rm gra}$ over $[1.0,6.0]$, giving 919,677 dust extinction-curve models.

\subsection{\texorpdfstring{Matching model curves and estimating dust mass surface density}{Matching model curves and estimating dust mass surface density}}

The observed inputs are the optical attenuation curve $A_{\rm opt}(\lambda)$ and the UV attenuation values $A_{w2}$, $A_{m2}$, and $A_{w1}$, while the model library provides $\kappa(\lambda)=\kappa_{\rm abs}(\lambda)+\kappa_{\rm sca}(\lambda)$. To compare curve shapes independently of dust mass surface density, we first normalize both by their V-band values:
\begin{equation}
{A(\lambda)}/{A_{V}}={\kappa(\lambda)}/{\kappa_{V}}.
\end{equation}
We then compare the normalized observed curve with each normalized model curve, compute separate $\chi^2$ values for the optical and NUV constraints, and define $\chi^2=\chi^2_{\rm opt}+\chi^2_{\rm NUV}$.

For each template, we then fit the original, unnormalized attenuation curve by treating $\Sigma_{\rm dust}$ as the amplitude parameter. Each template is assigned a weight $\exp[-(\chi^2-\chi^2_{\min})/2]$, and the weighted-average $\Sigma_{\rm dust}$ is adopted as the dust mass surface-density estimate. For the selected sample, the weighted-average and highest-weight values of $q_{\rm sil}$, $a_{\max}$, $\alpha_{\rm sil}$, and $\alpha_{\rm gra}$ are very similar, with Pearson correlation coefficients of 0.998, 0.998, 0.990, and 0.988, respectively. The 84th-percentile absolute differences are $4.3\times10^{-8}$ in $q_{\rm sil}$, $3.1\times10^{-4}\,\mu{\rm m}$ in $a_{\max}$, 0.0014 in $\alpha_{\rm sil}$, and 0.0010 in $\alpha_{\rm gra}$. We therefore use the highest-weight template as the best-matched model and adopt its parameters for the dust-property analysis.

Some degeneracy is unavoidable: similar normalized attenuation curves can be produced by different combinations of $a_{\max}$, $\alpha_{\rm sil}$, $\alpha_{\rm gra}$, and $q_{\rm sil}$, and the conversion from attenuation amplitude to $\Sigma_{\rm dust}$ depends on the assumed foreground-screen geometry. The comparison above shows that the main parameter assignments are not sensitive to using weighted averages instead of highest-weight templates, but it does not remove the intrinsic model degeneracy. We therefore emphasize relative trends among spaxels and the mapping between observed curve features and effective model parameters, rather than the uniqueness of any individual grain-size distribution or dust mass.

\begin{figure}[!t]
\centering
\includegraphics[width=\columnwidth]{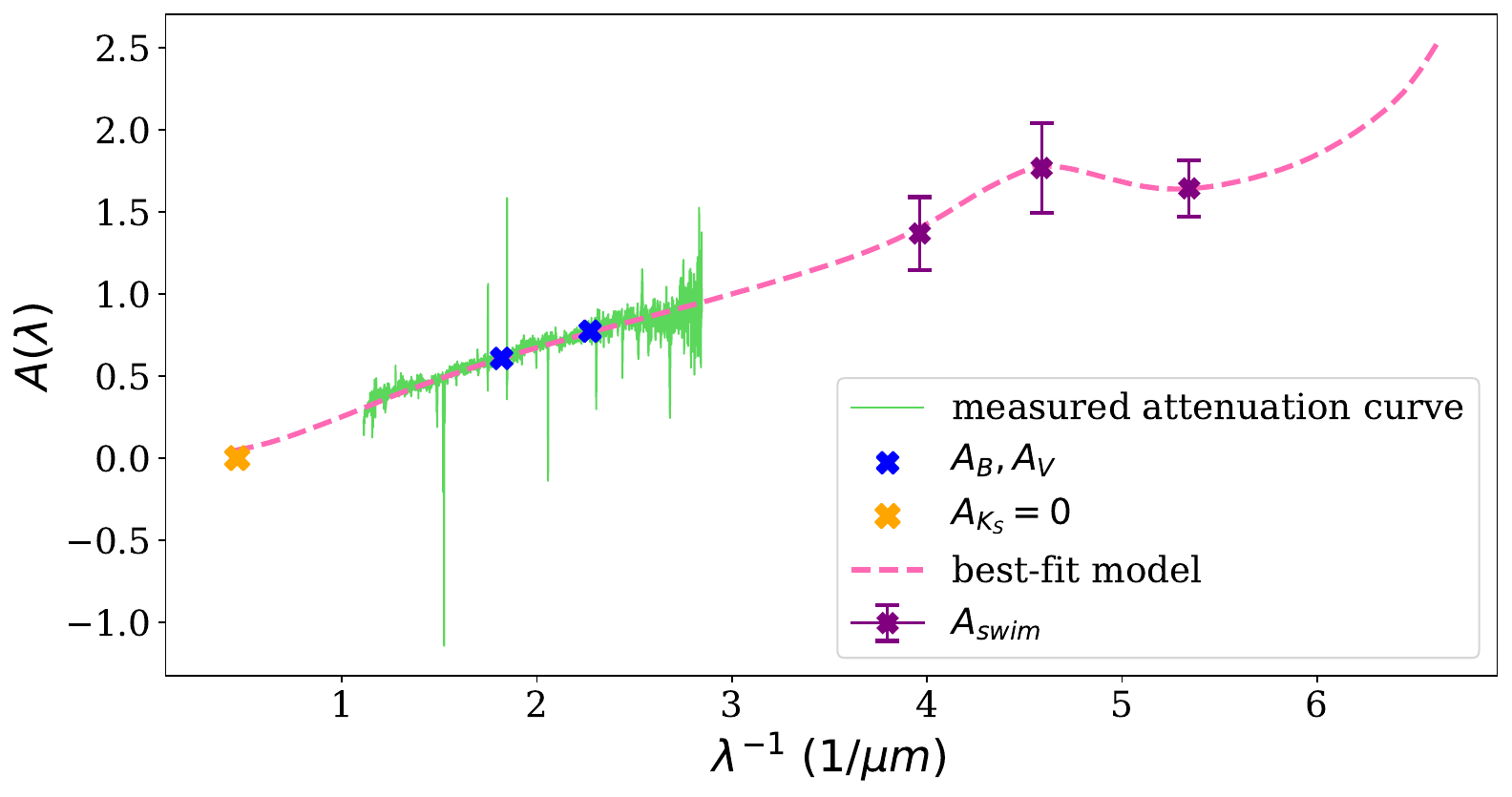}
\caption{Example of the dust-model fit for one spaxel in galaxy 8138-12704. The green curve shows the optical attenuation curve measured from the MaNGA spectrum. Blue crosses mark the derived $B$- and $V$-band attenuations, and purple points show the UVOT-band attenuations $A_{w2}$, $A_{m2}$, and $A_{w1}$, with uncertainties from error propagation. The orange cross marks the adopted $K_s$-band anchor, where the attenuation is assumed to be negligible. The pink dashed curve is the best-matched dust-model attenuation curve.\label{fig:fit_example}}
\end{figure}

\begin{figure}[!htbp]
        \centering
        \includegraphics[width=\columnwidth]{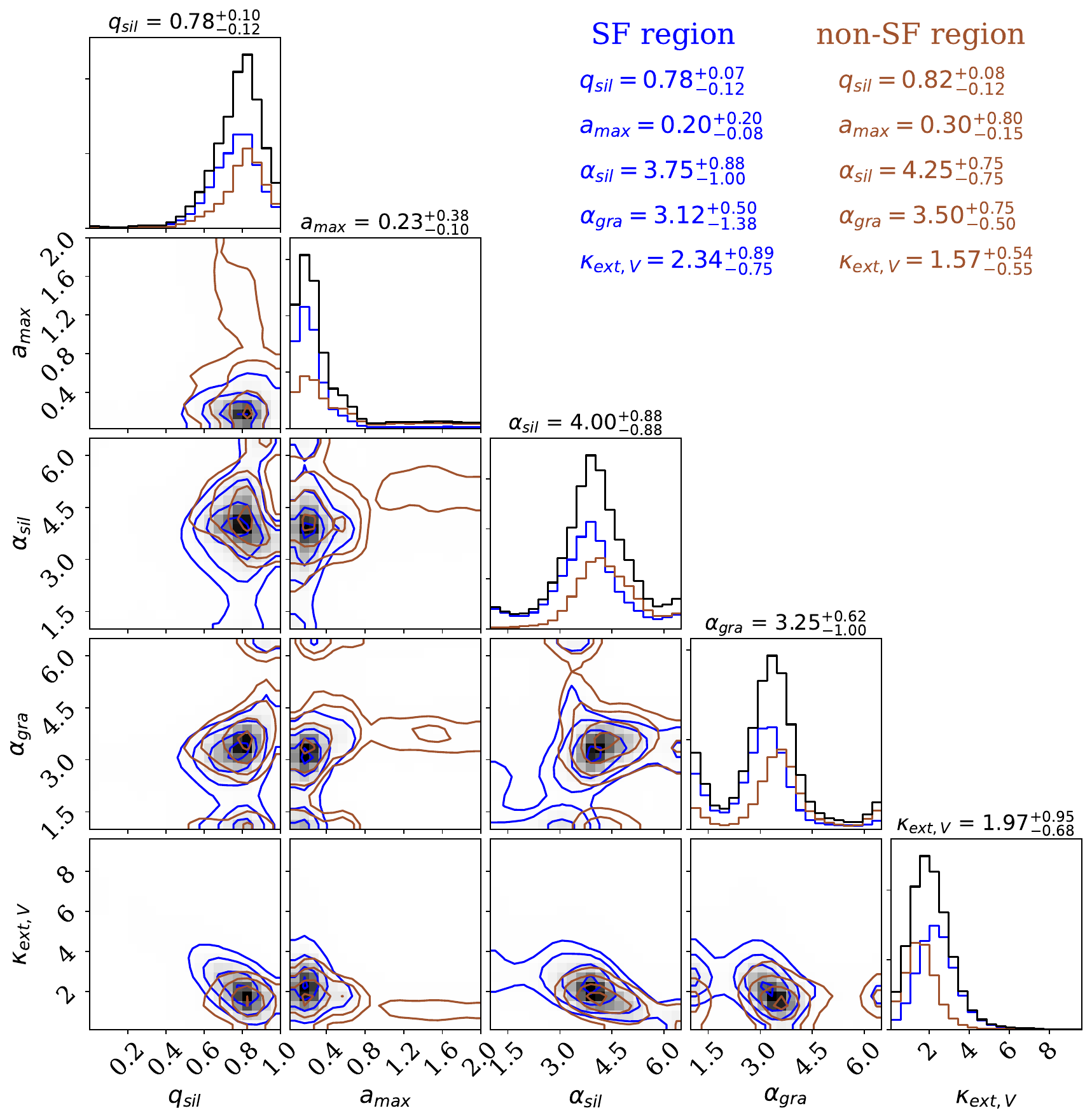}
        \caption{Joint distributions of the best-matched dust-model parameters. Black shows the full sample, while blue and sienna contours show SF and non-SF regions. The one-dimensional panels give the corresponding marginalized distributions and quantiles.\label{fig:param_posterior}}
\end{figure}

\autoref{fig:fit_example} illustrates one representative fit. The best-matched model reproduces the broad optical-to-NUV attenuation shape, showing how the measured curve maps onto a dust composition, size distribution, and dust mass surface density. \autoref{fig:param_posterior} summarizes the best-matched parameters for the full, SF, and non-SF samples. Non-SF regions favor a higher silicate mass fraction and extend to larger maximum grain sizes, while some fitted parameters lie close to the grid edge; these trends should therefore be interpreted as effective, model-dependent results.

\subsection{Derived dust properties} 

From the model matching we derive five dust-related quantities: dust mass surface density, dust-to-stellar mass ratio, silicate mass fraction, small-silicate mass fraction, and small-graphite mass fraction. The dust mass surface density $\Sigma_{\rm dust}$ comes from the attenuation-amplitude fit and is corrected for host-galaxy inclination. We define the dust-to-stellar mass ratio as $\log(\Sigma_{\rm dust}/\Sigma_\ast)$, and the silicate mass fraction is the best-matched model parameter $q_{\rm sil}$.

To describe the relative abundance of small grains, we integrate each fitted MRN-like distribution and compute the mass in grains with $a<0.01\,\mu{\rm m}$, a scale relevant for UV attenuation and the 2175\AA\ feature. This value is used as a diagnostic threshold for very small grains rather than as a unique physical boundary. The small-grain masses in the silicate and graphite components are denoted $m_{a<0.01,{\rm sil}}$ and $m_{a<0.01,{\rm gra}}$, and we use the total-dust-mass fractions $\log_{10}(m_{a<0.01,{\rm sil}}/m_{\rm Tot})$ and $\log_{10}(m_{a<0.01,{\rm gra}}/m_{\rm Tot})$.

\begin{figure*}[!t]
        \begin{minipage}{0.2\textwidth}
        \gridline{\fig{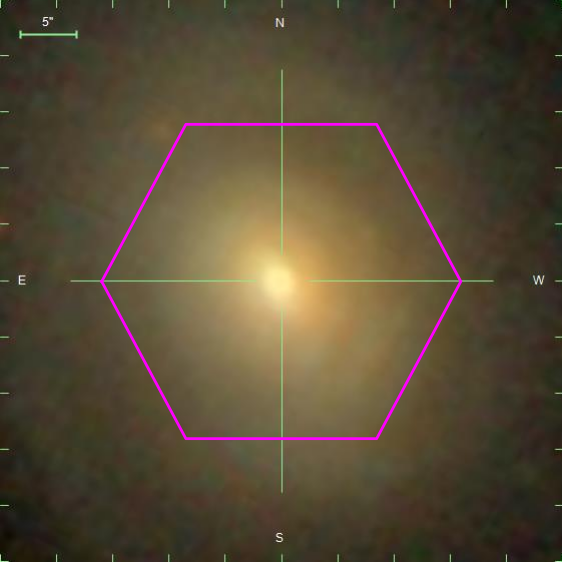}{0.8\textwidth}{}}
    \end{minipage}
    \begin{minipage}{0.8\textwidth}
        \gridline{\fig{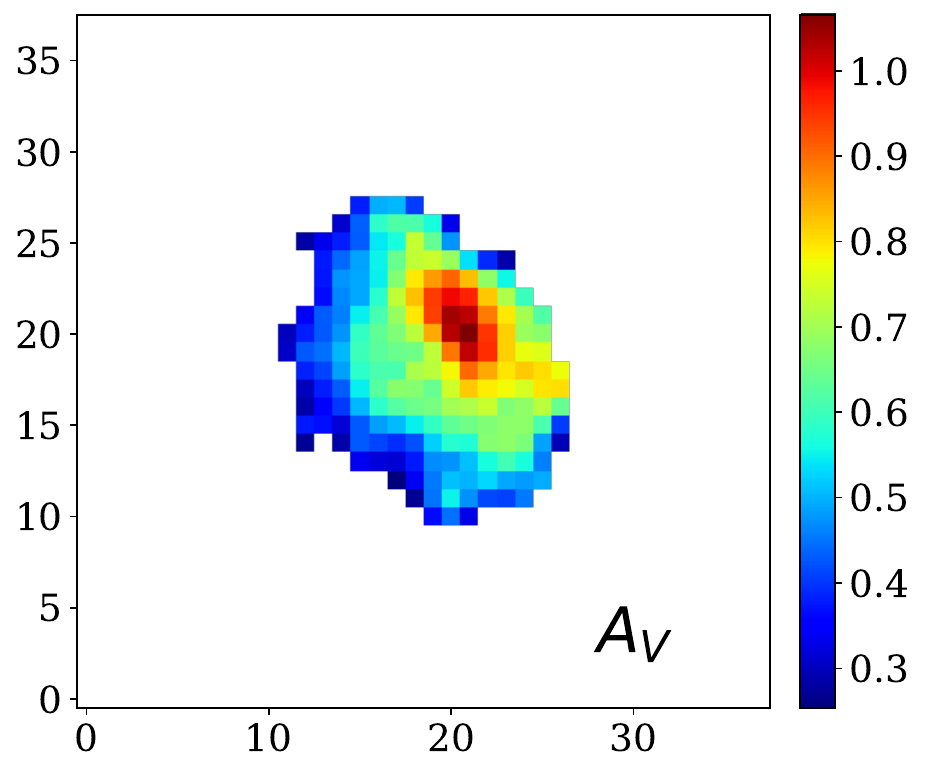}{0.25\textwidth}{}
                \fig{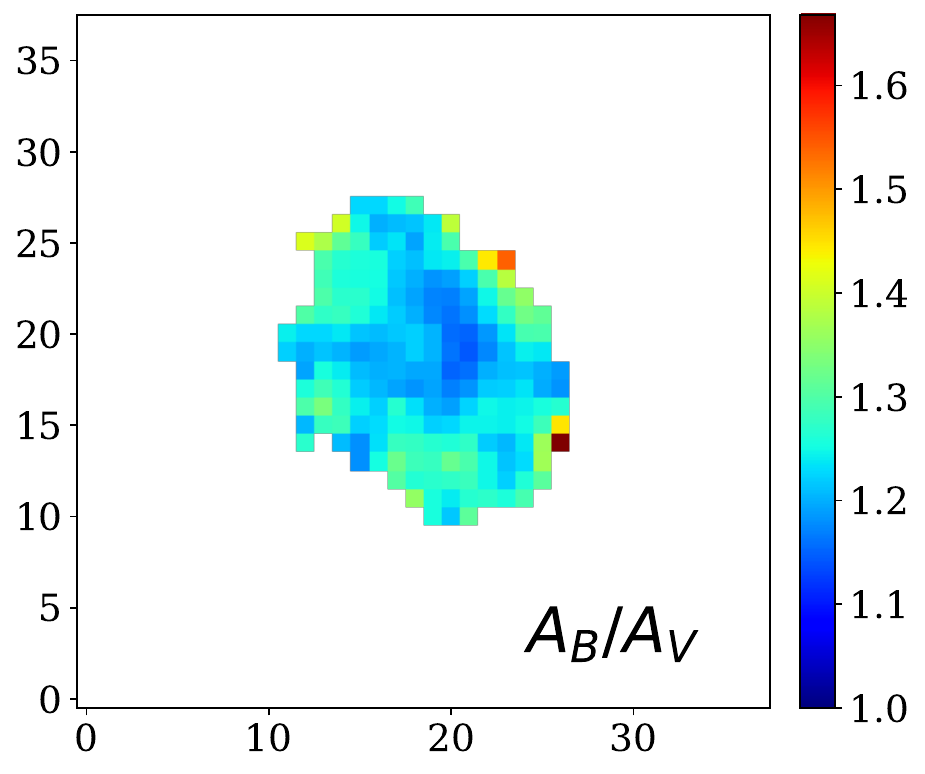}{0.25\textwidth}{}
                \fig{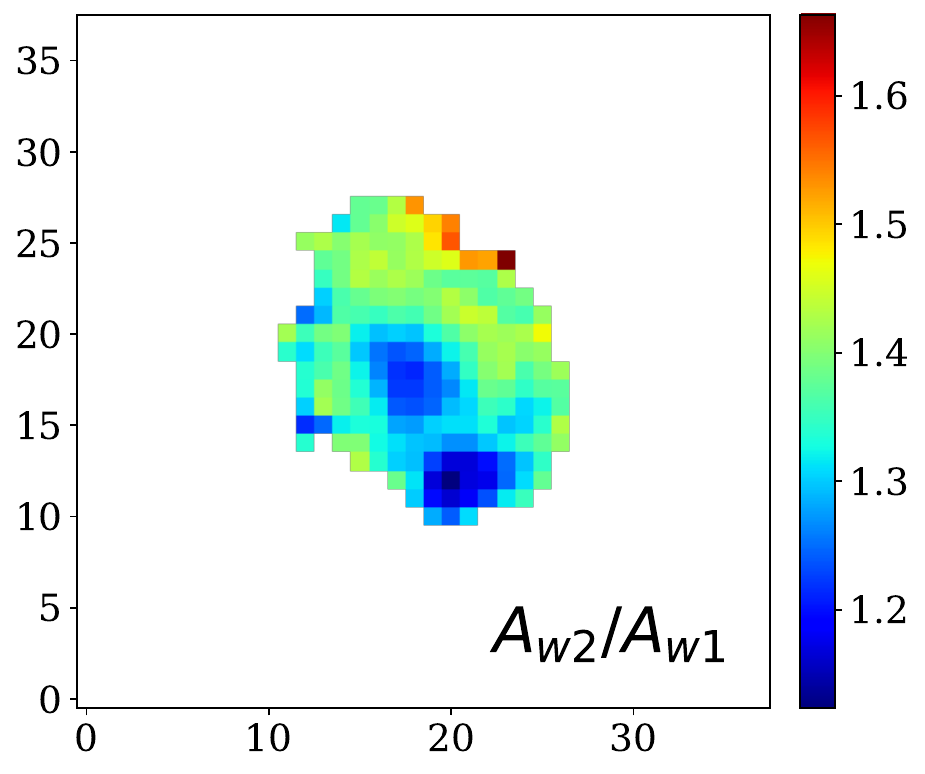}{0.25\textwidth}{}
                \fig{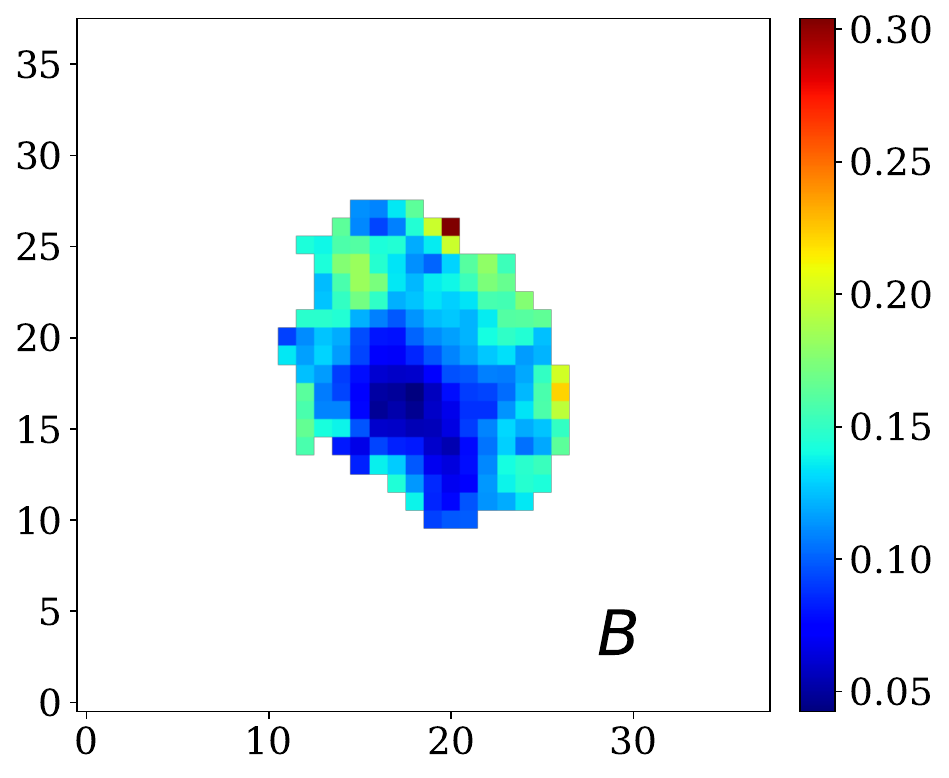}{0.25\textwidth}{}}
    \end{minipage}
    \begin{minipage}{1.0\textwidth}
        \gridline{\fig{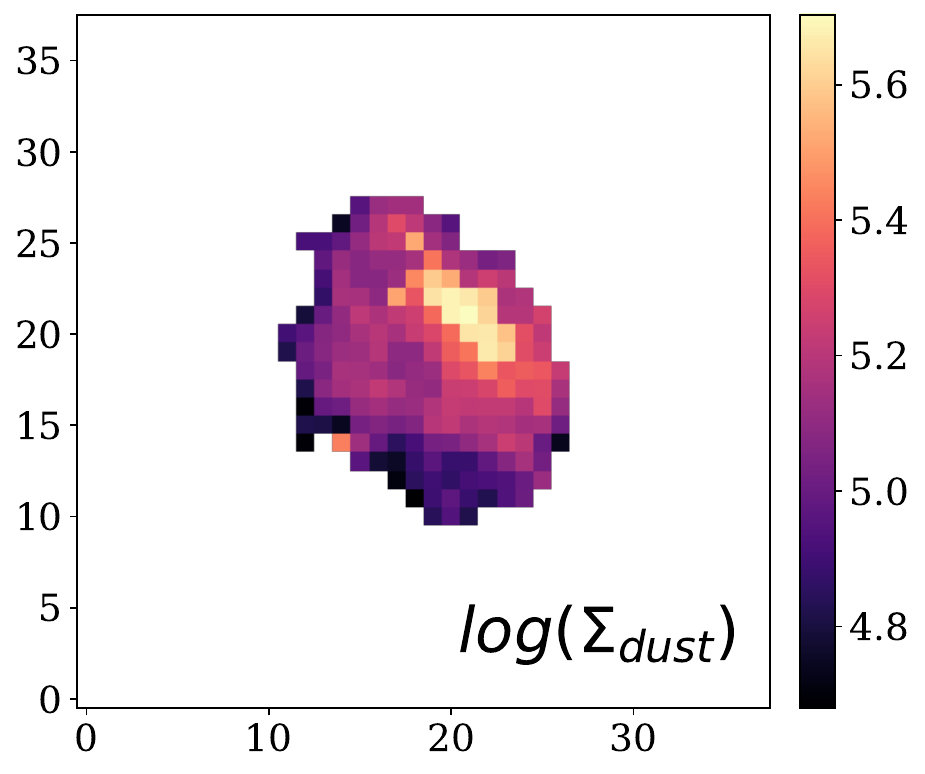}{0.2\textwidth}{}
                \fig{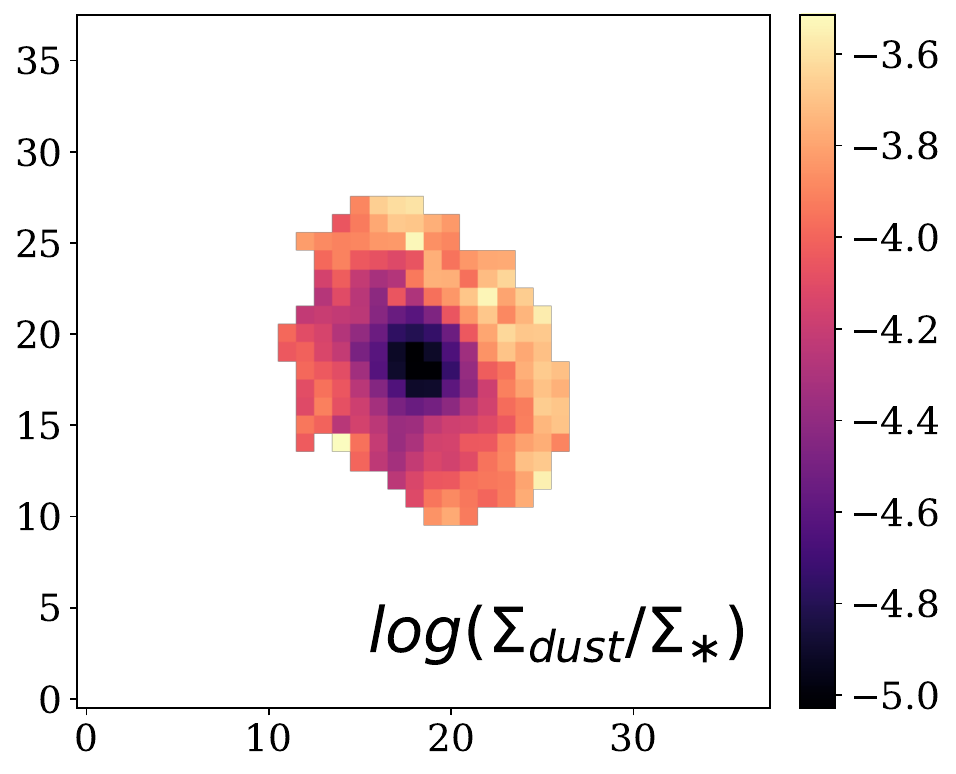}{0.2\textwidth}{}
                \fig{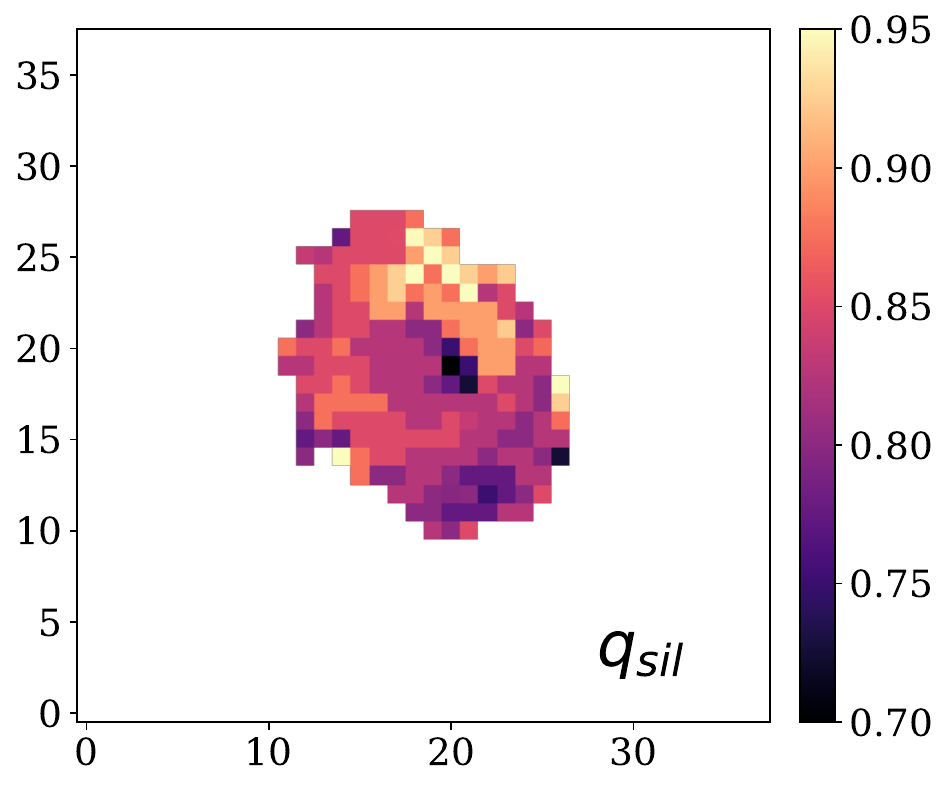}{0.2\textwidth}{}
                \fig{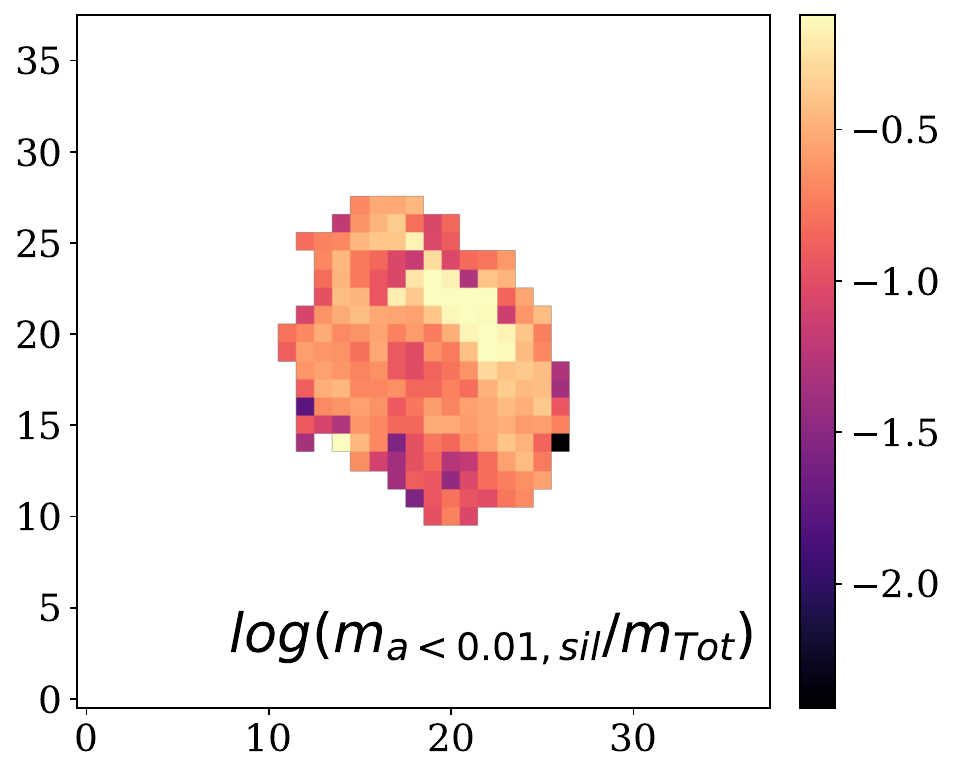}{0.2\textwidth}{}
                \fig{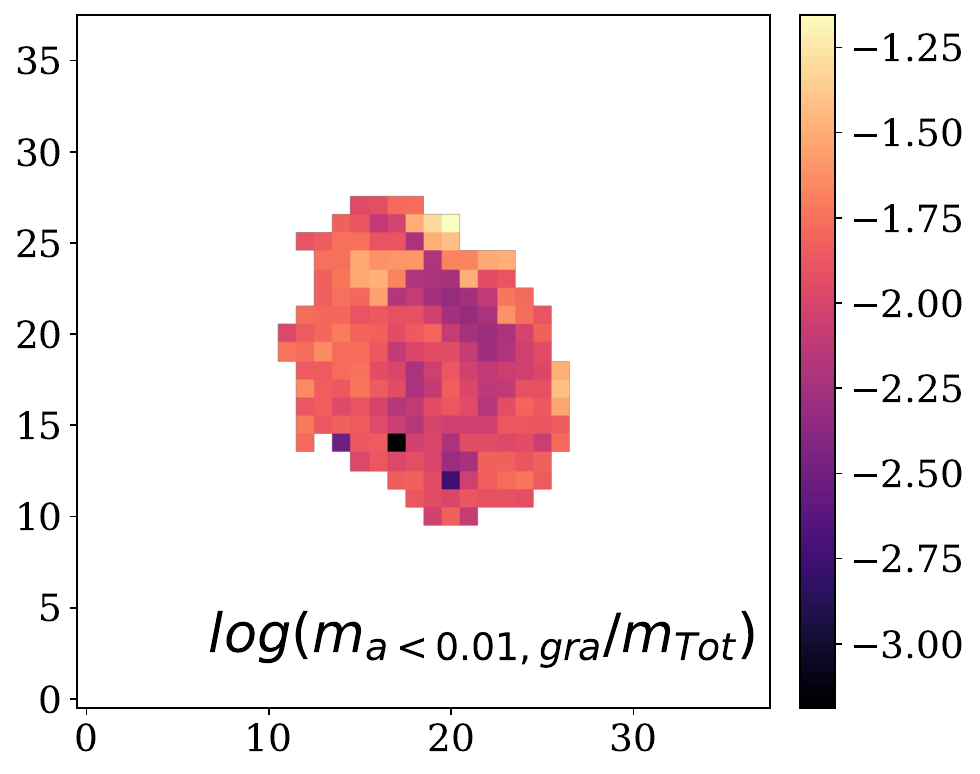}{0.2\textwidth}{}}
    \end{minipage}
        \caption{Example attenuation and dust-property maps for galaxy 11748-12705. The first row shows, from left to right, the SDSS optical image, $A_V$, $A_B/A_V$, $A_{w2}/A_{w1}$, and the attenuation-curve relative 2175\AA\ bump strength $B$. The second row shows the model-inferred dust mass surface density, dust-to-stellar mass ratio, silicate mass fraction, small-silicate mass fraction, and small-graphite mass fraction.\label{fig:example1_DM}}
\end{figure*}

\autoref{fig:example1_DM} compares the measured attenuation properties and derived dust properties for one representative galaxy. The maps show coherent spatial variations across the MaNGA footprint and illustrate that the attenuation-curve-based estimates use both curve amplitude and shape, rather than a simple conversion from $A_V$ to dust mass.

\begin{figure}[!t]
    \gridline{\fig{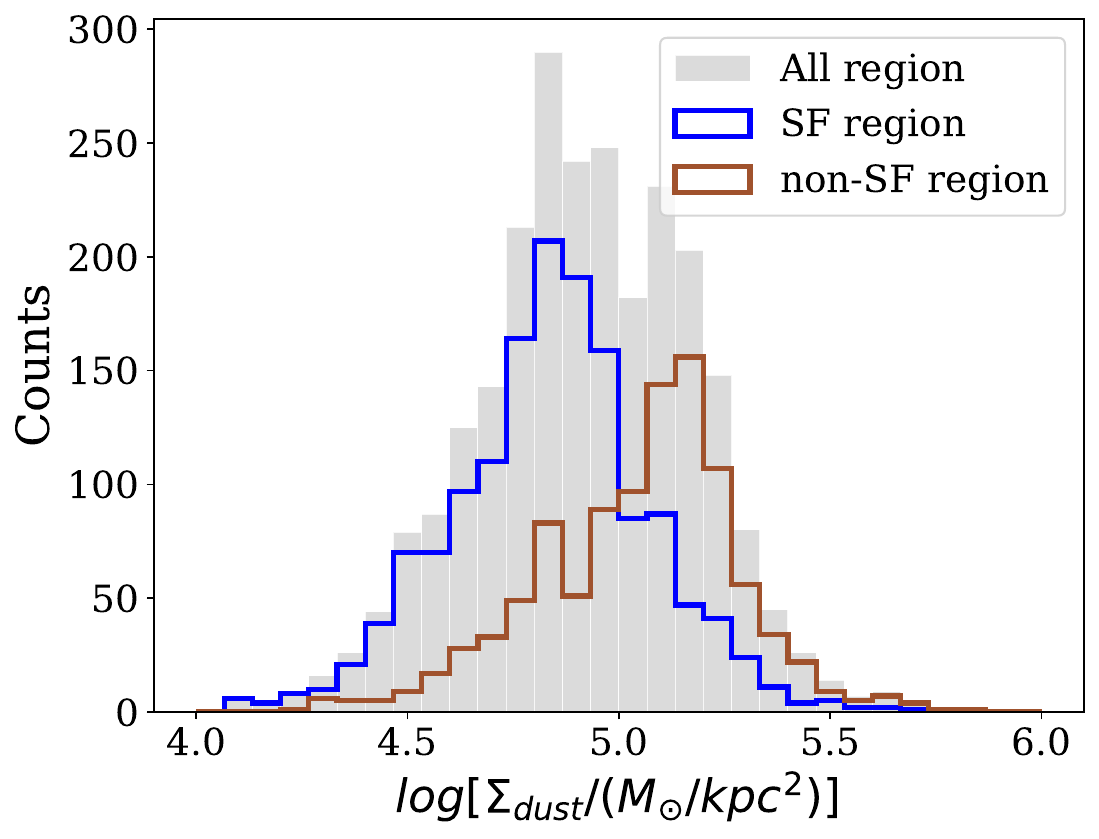}{0.49\columnwidth}{(a)}
        \fig{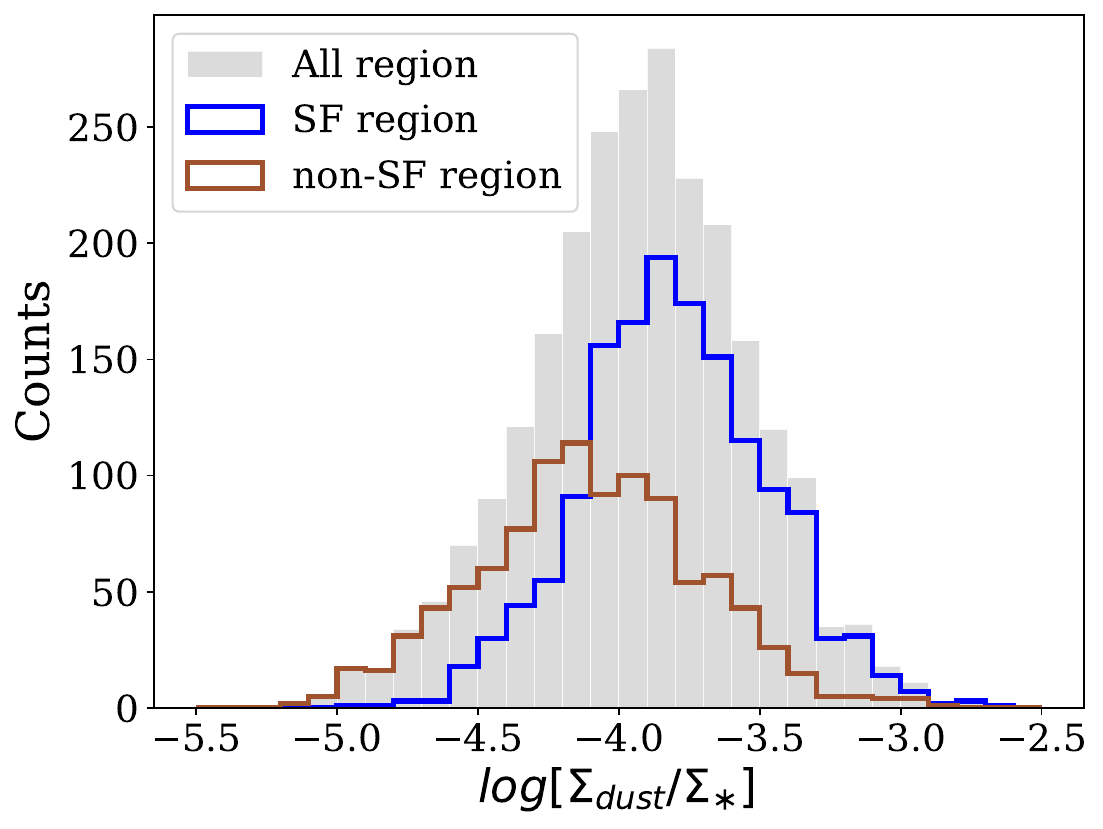}{0.49\columnwidth}{(b)}}
    \gridline{\fig{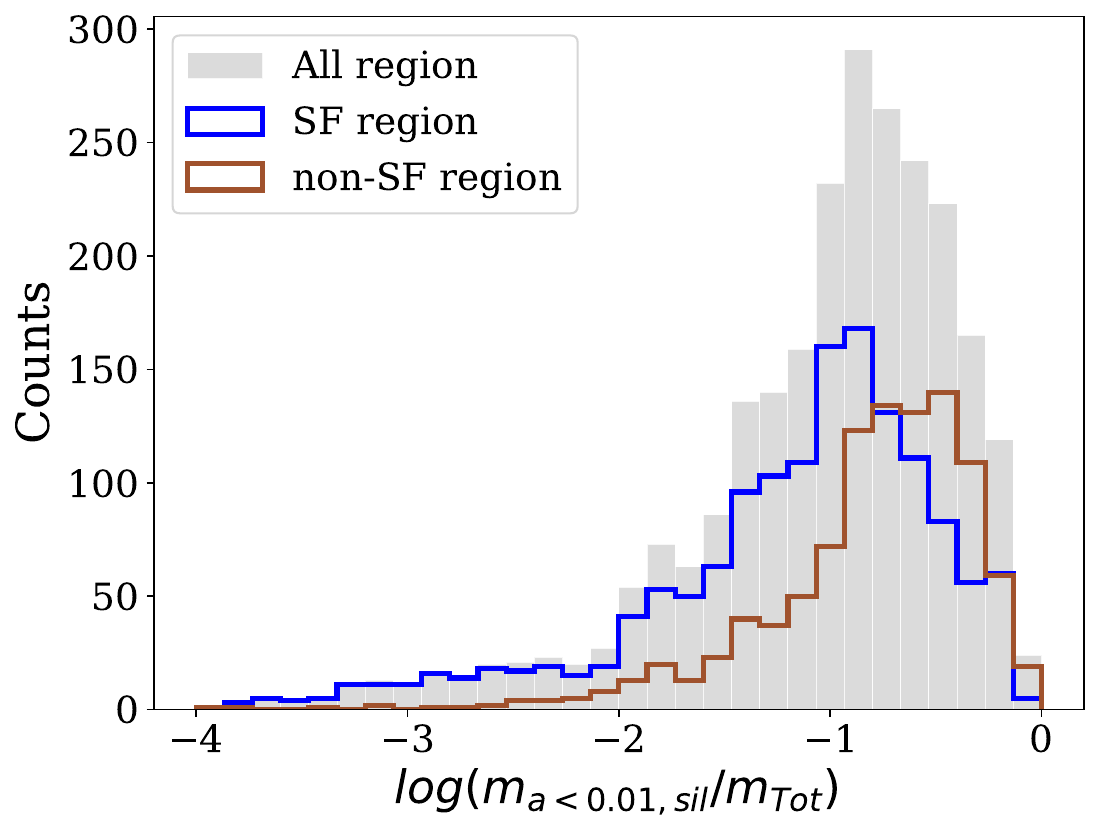}{0.49\columnwidth}{(c)}
        \fig{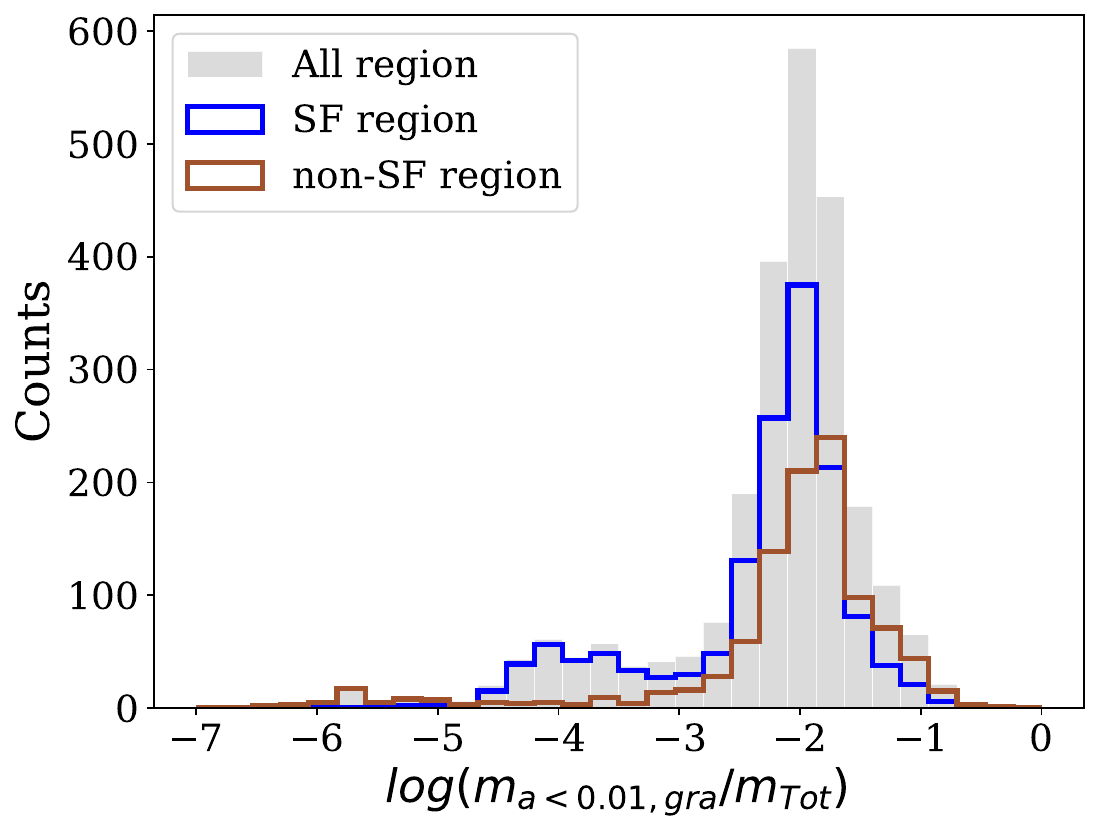}{0.49\columnwidth}{(d)}}
	\caption{Distributions of derived dust quantities: dust mass surface density, dust-to-stellar mass ratio, small-silicate mass fraction, and small-graphite mass fraction. Gray histograms show the full selected sample, while blue and sienna step histograms show SF and non-SF regions. Non-SF regions have larger dust mass surface densities, lower dust-to-stellar mass ratios, and higher inferred silicate and small-grain fractions than SF regions.\label{fig:derived_dust_distribution}}
\end{figure}

\autoref{fig:derived_dust_distribution} gives the sample-wide distributions. The dust mass surface density is concentrated around $10^4$--$10^6\,M_\odot\,{\rm kpc}^{-2}$. Non-SF regions have larger $\Sigma_{\rm dust}$, while SF regions have larger $\Sigma_{\rm dust}/\Sigma_\ast$, showing that the higher absolute dust content in non-SF regions is largely associated with their larger stellar mass surface densities. Non-SF regions also have higher inferred small-grain fractions, while some SF spaxels have very low small-graphite fractions where $\alpha_{\rm gra}$ approaches the edge of the model grid.

\section{Results}\label{sec:results}

\subsection{\texorpdfstring{Connections between attenuation features and dust-model properties}{Connections between attenuation features and dust-model properties}}

\begin{figure*}[!thbp]
	\gridline{\fig{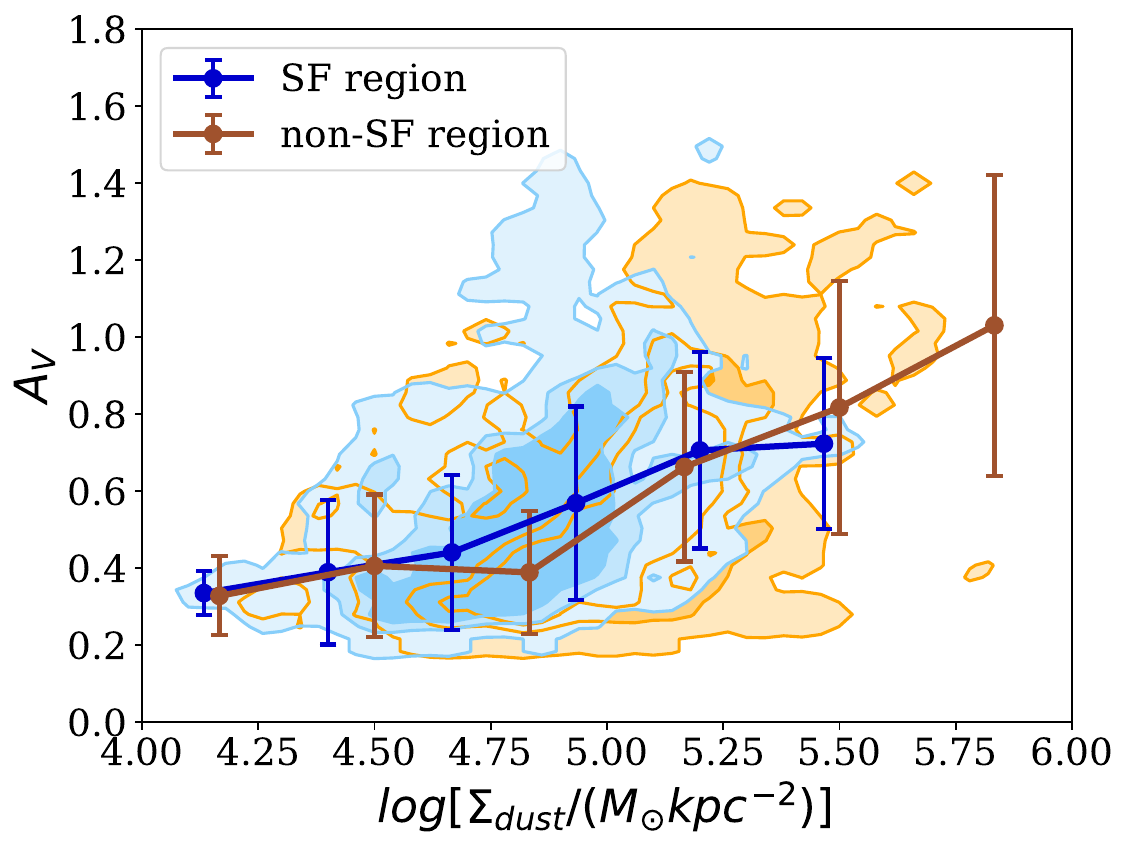}{0.245\textwidth}{}
		\fig{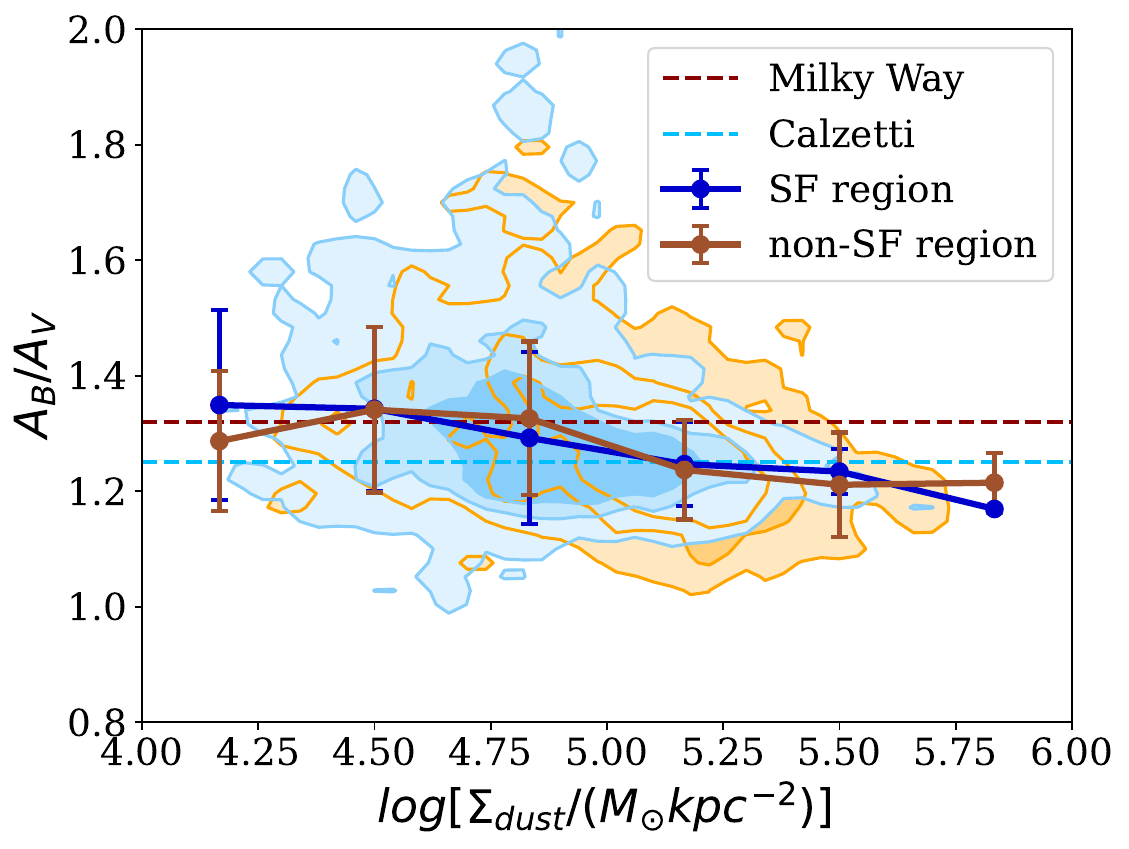}{0.245\textwidth}{}
	    \fig{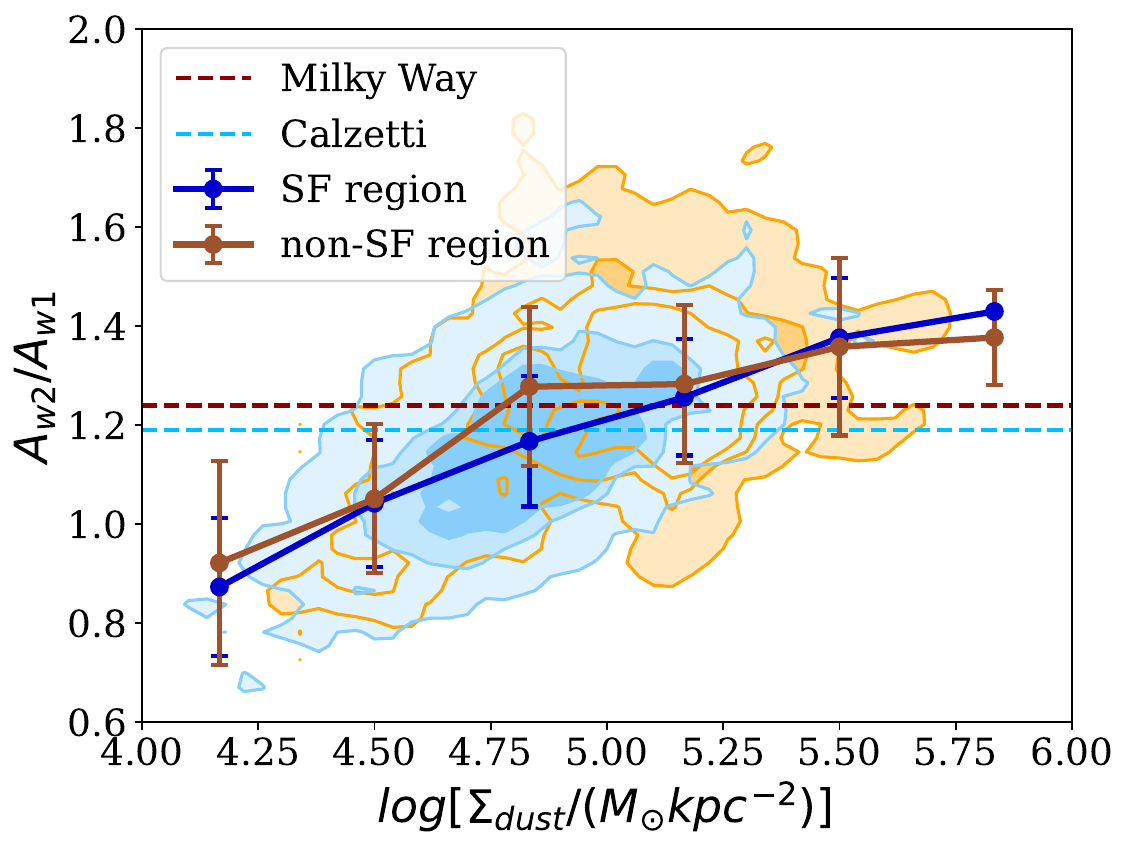}{0.245\textwidth}{}
   		\fig{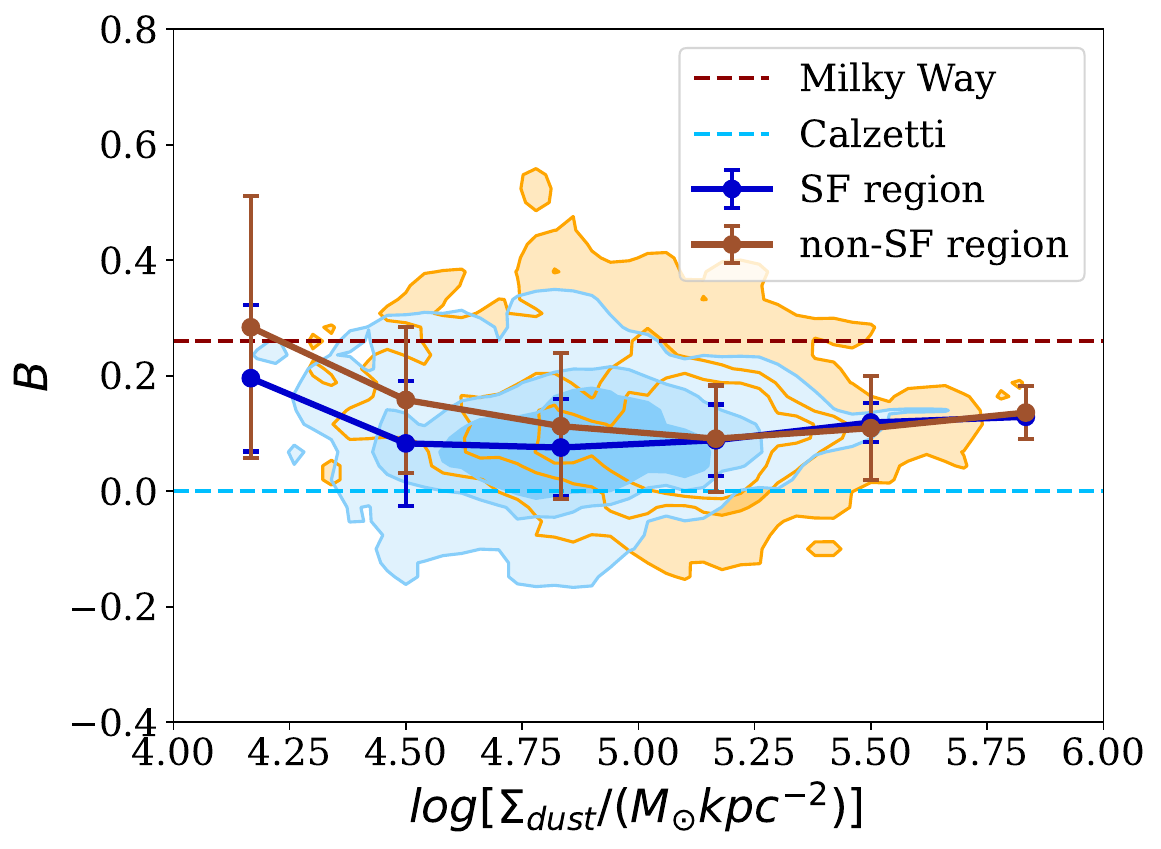}{0.245\textwidth}{}}
    \vspace{-1.6\baselineskip}
	\gridline{\fig{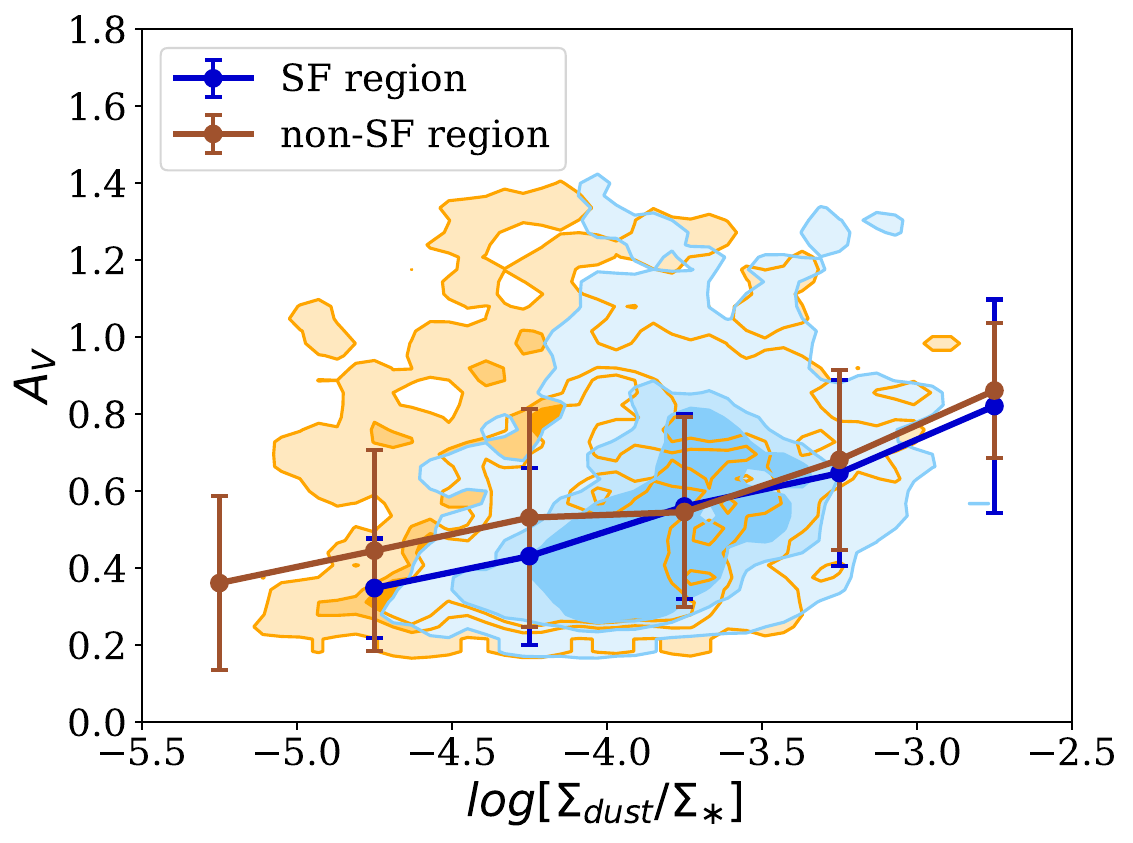}{0.245\textwidth}{}
		\fig{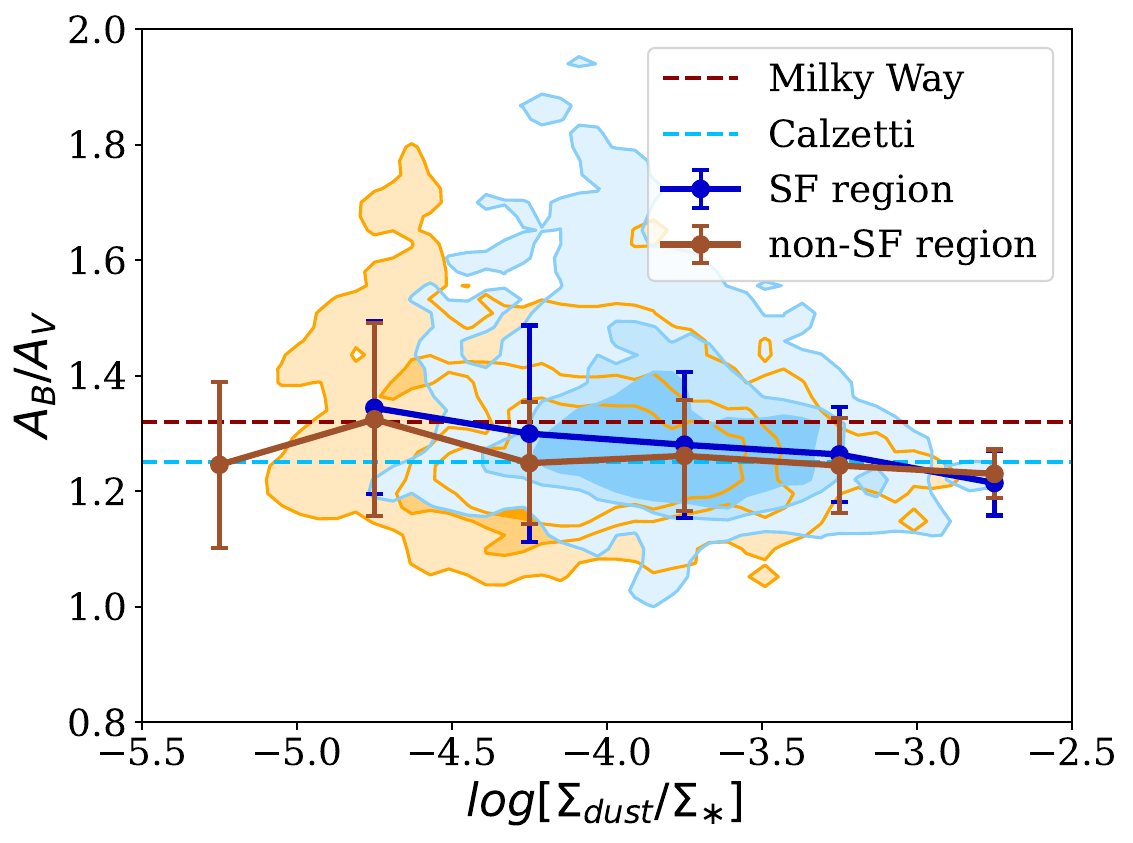}{0.245\textwidth}{}
		\fig{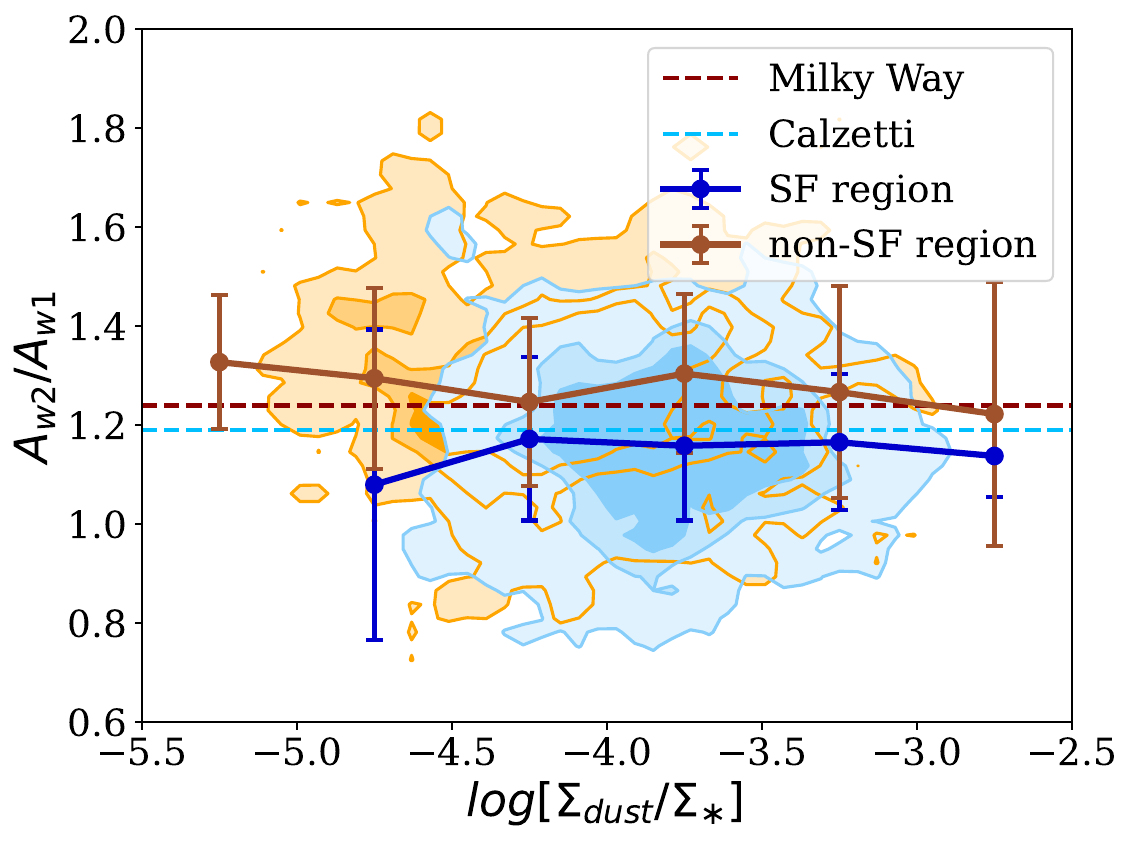}{0.245\textwidth}{}
		\fig{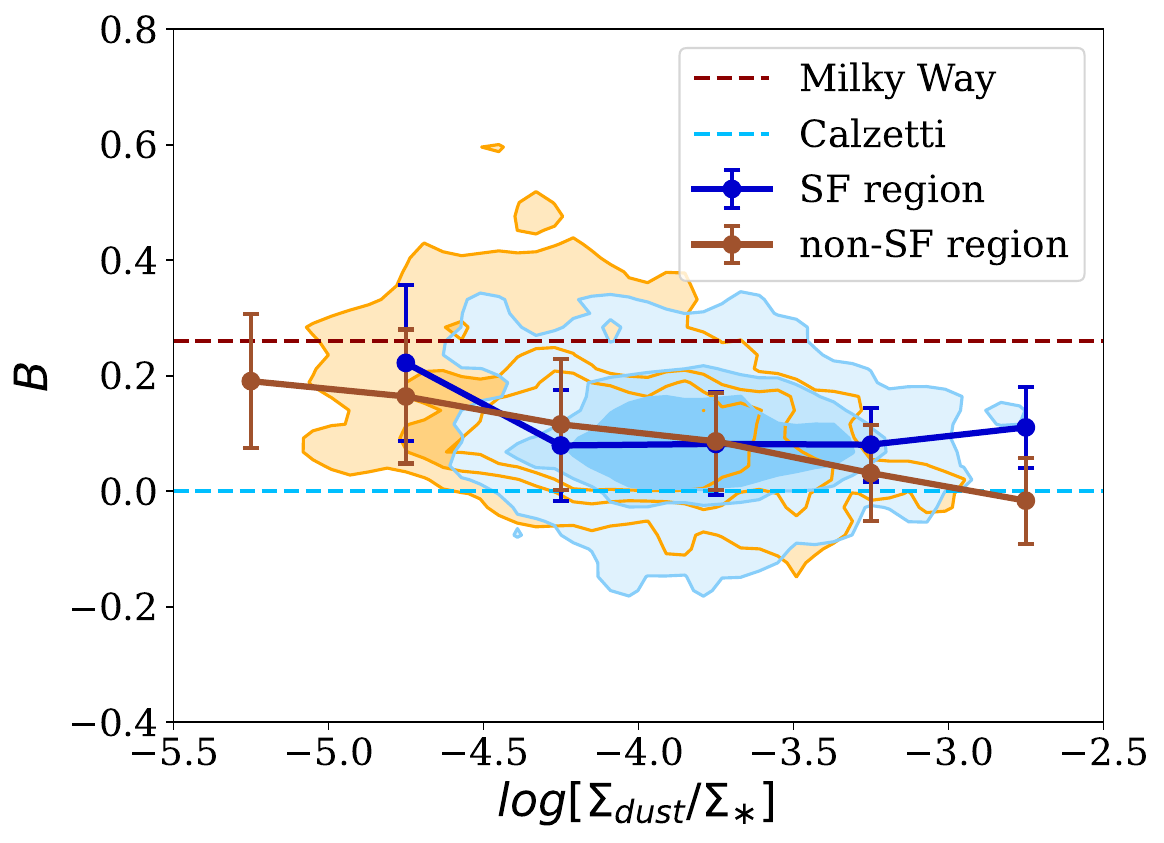}{0.245\textwidth}{}}
    \vspace{-1.6\baselineskip}
	\gridline{\fig{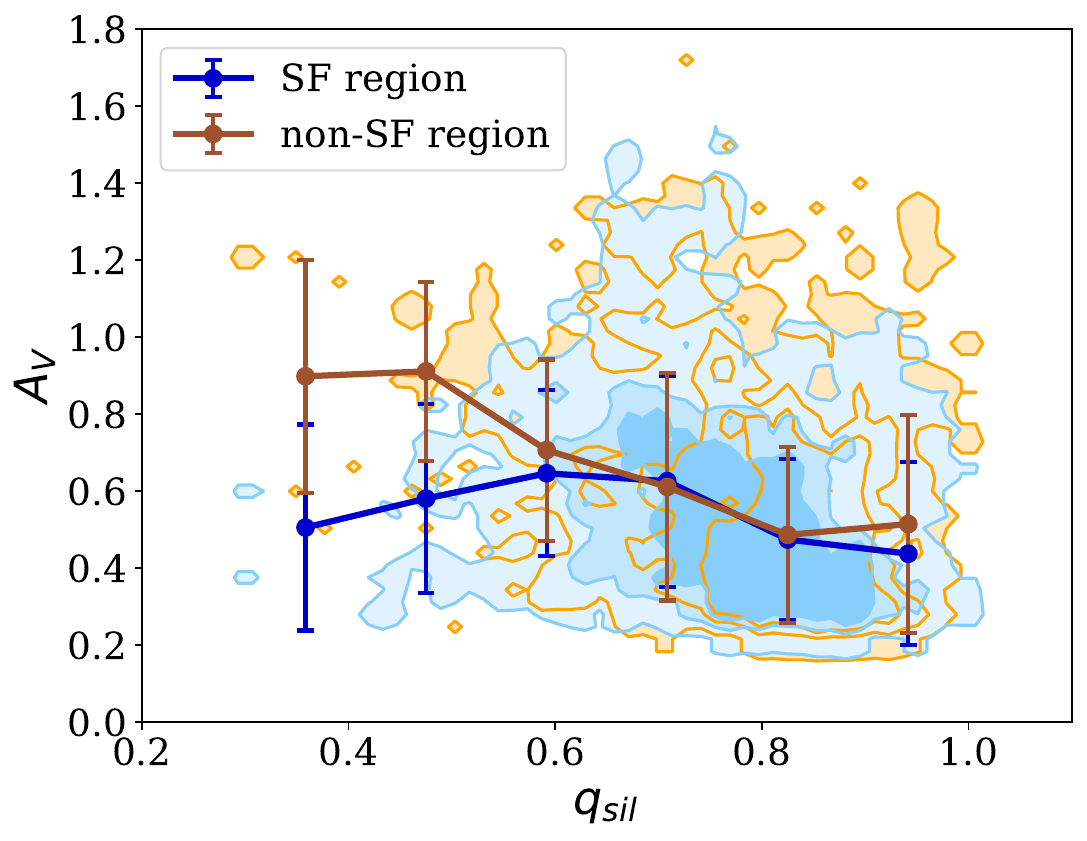}{0.245\textwidth}{}
		\fig{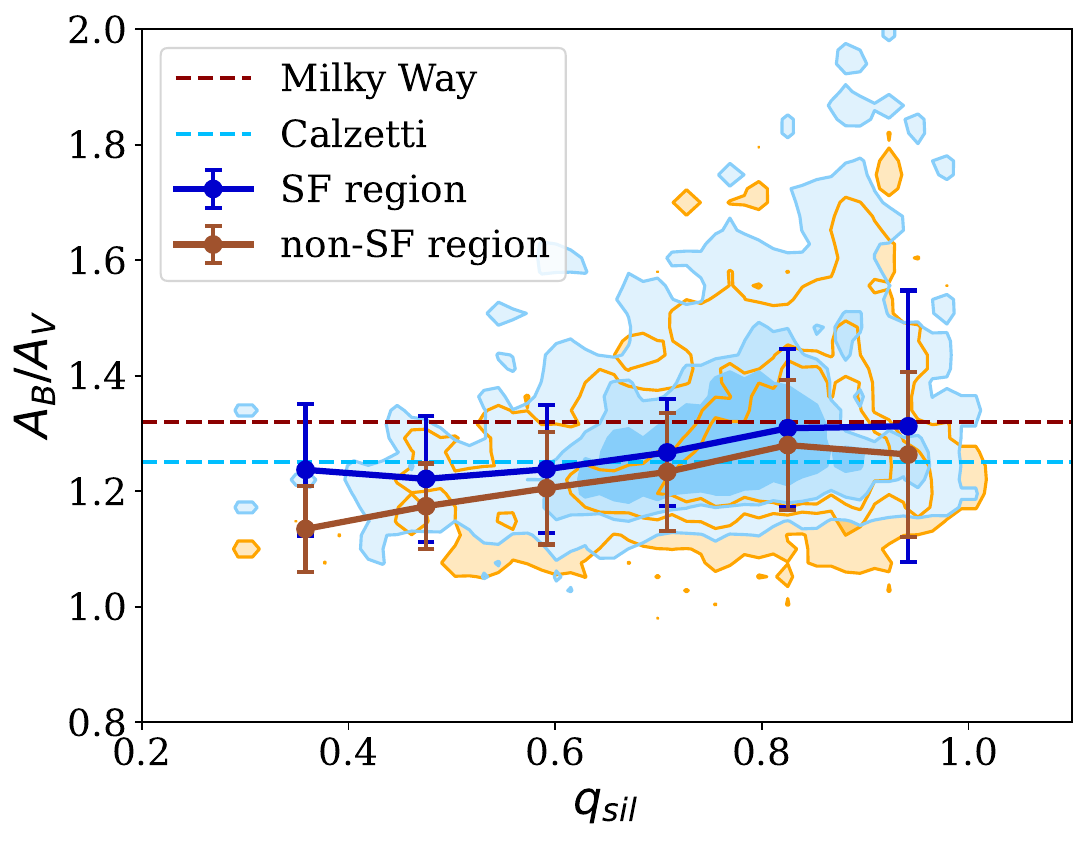}{0.245\textwidth}{}
		\fig{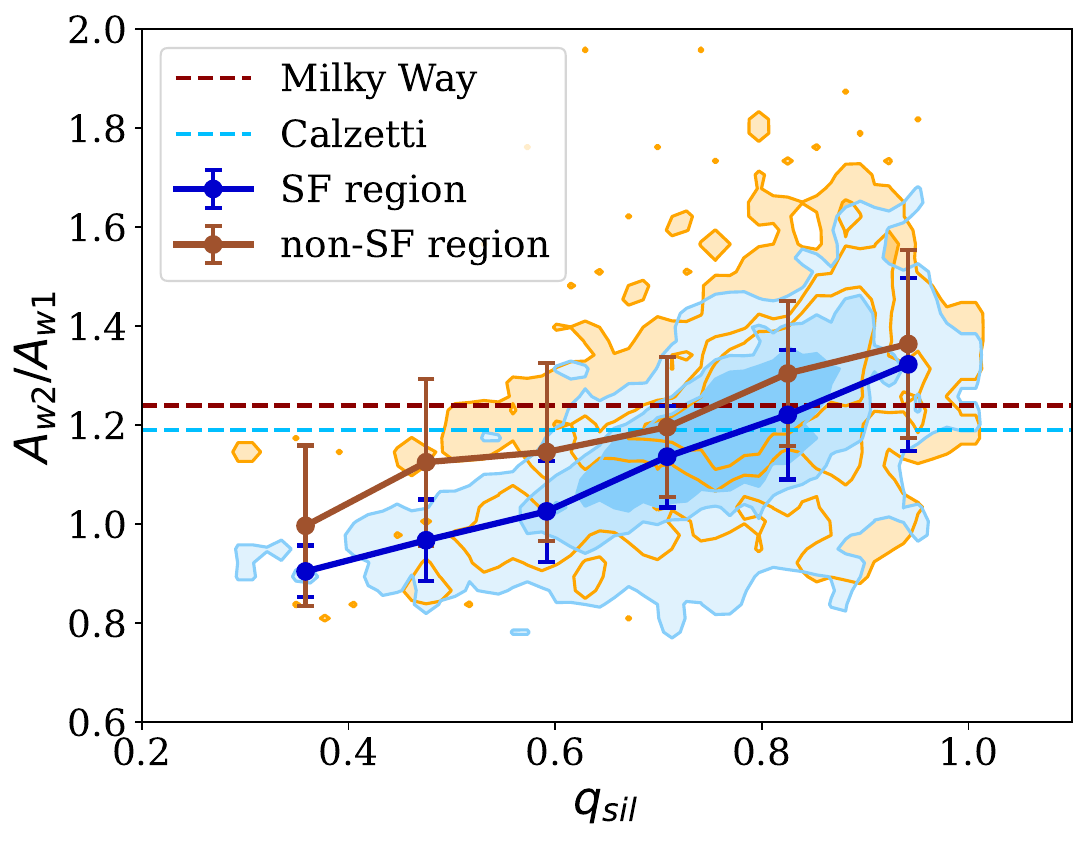}{0.245\textwidth}{}
		\fig{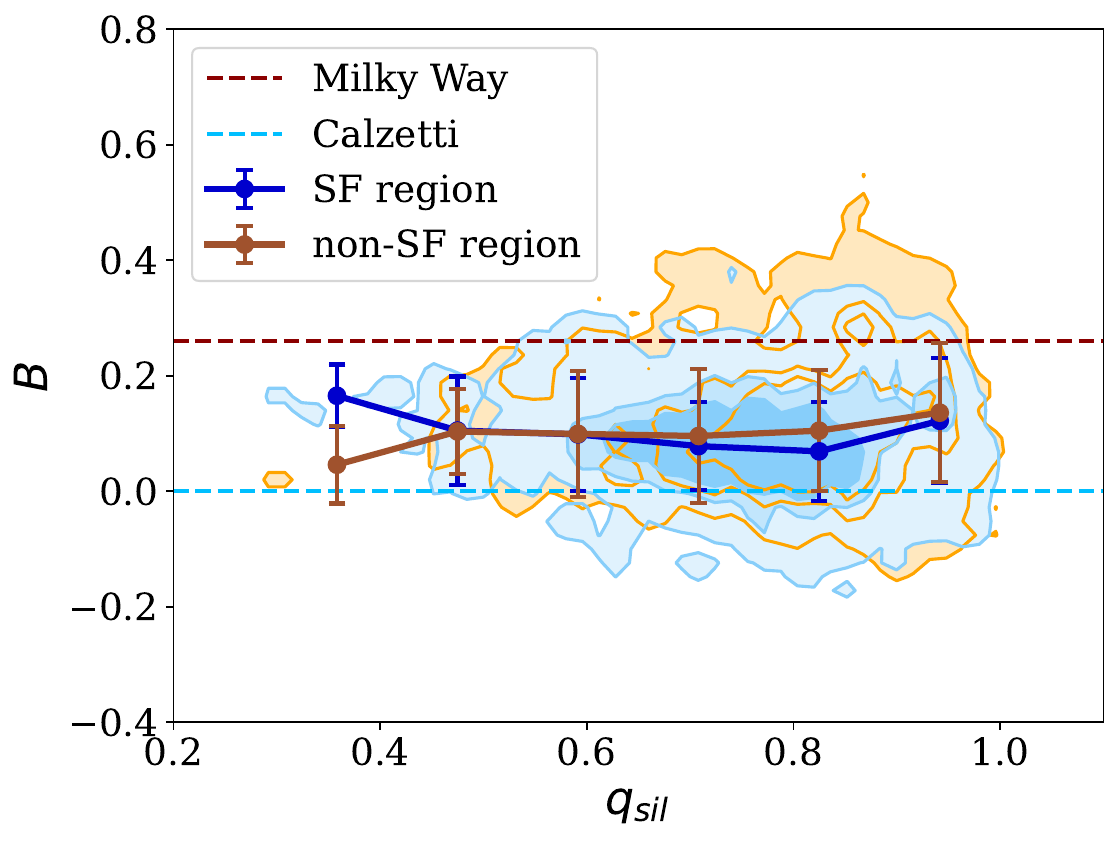}{0.245\textwidth}{}}
    \vspace{-1.6\baselineskip}
	\gridline{\fig{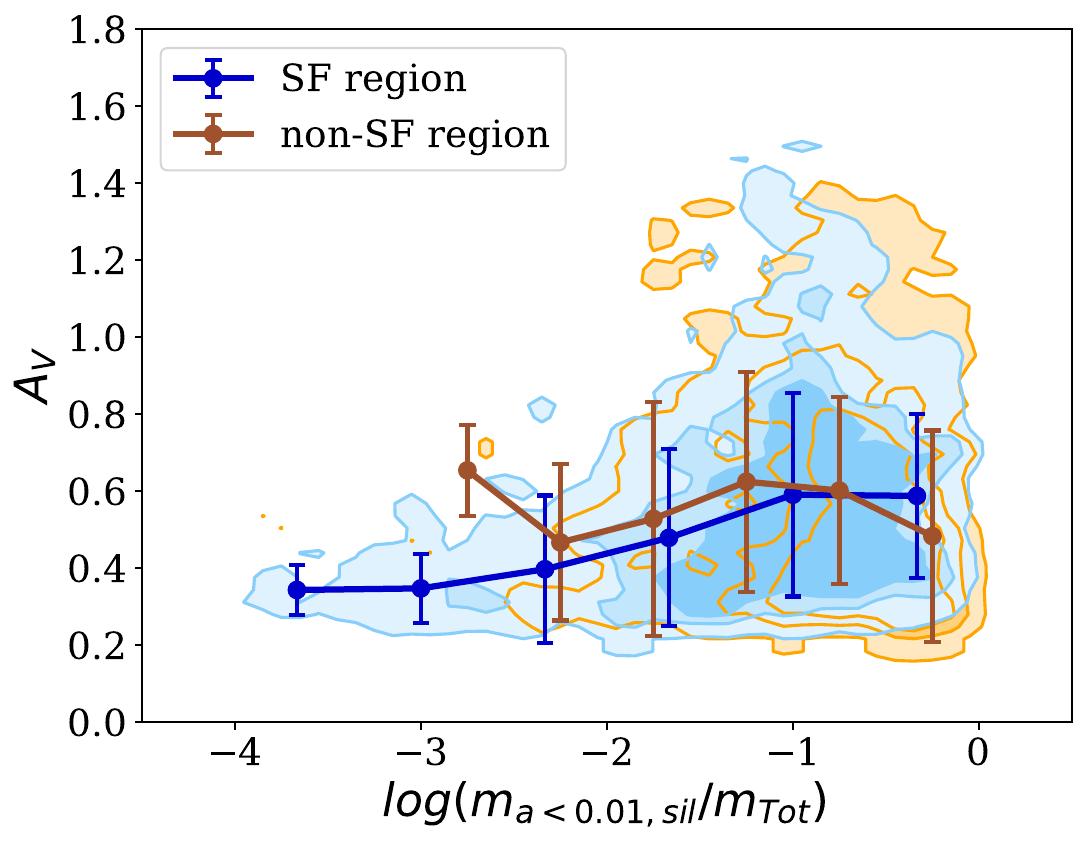}{0.245\textwidth}{}
		\fig{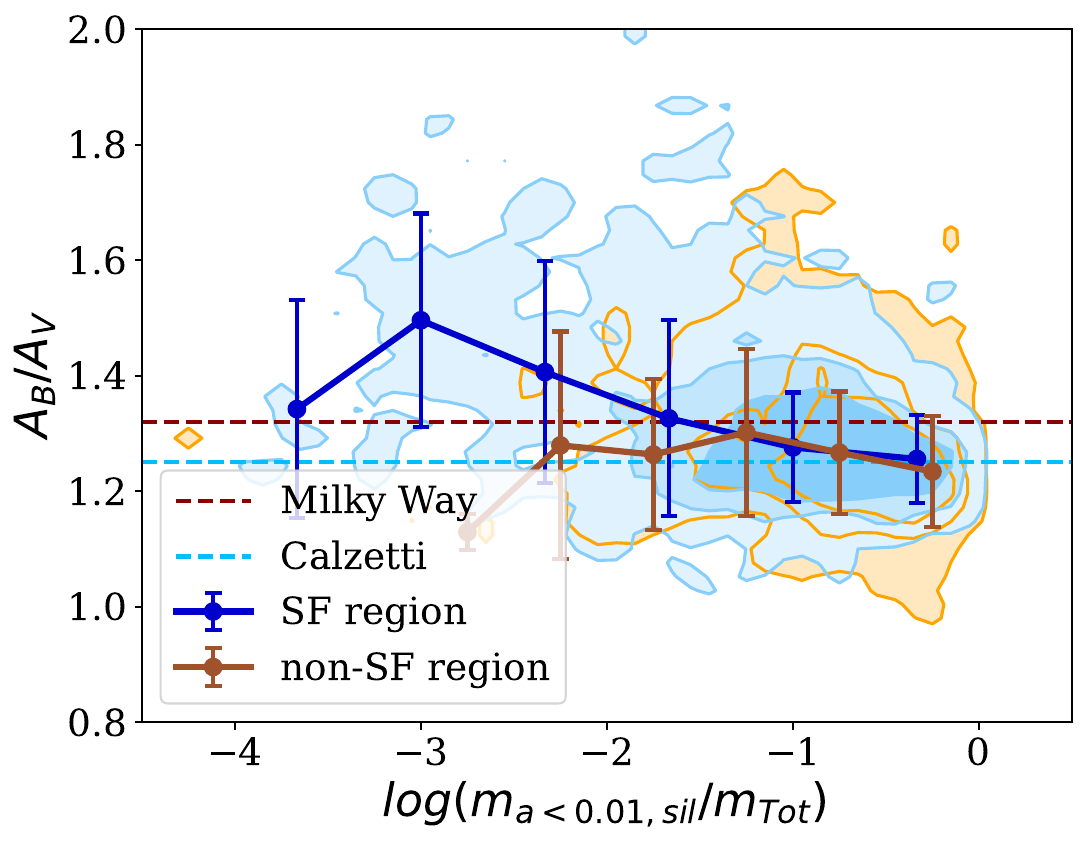}{0.245\textwidth}{}
		\fig{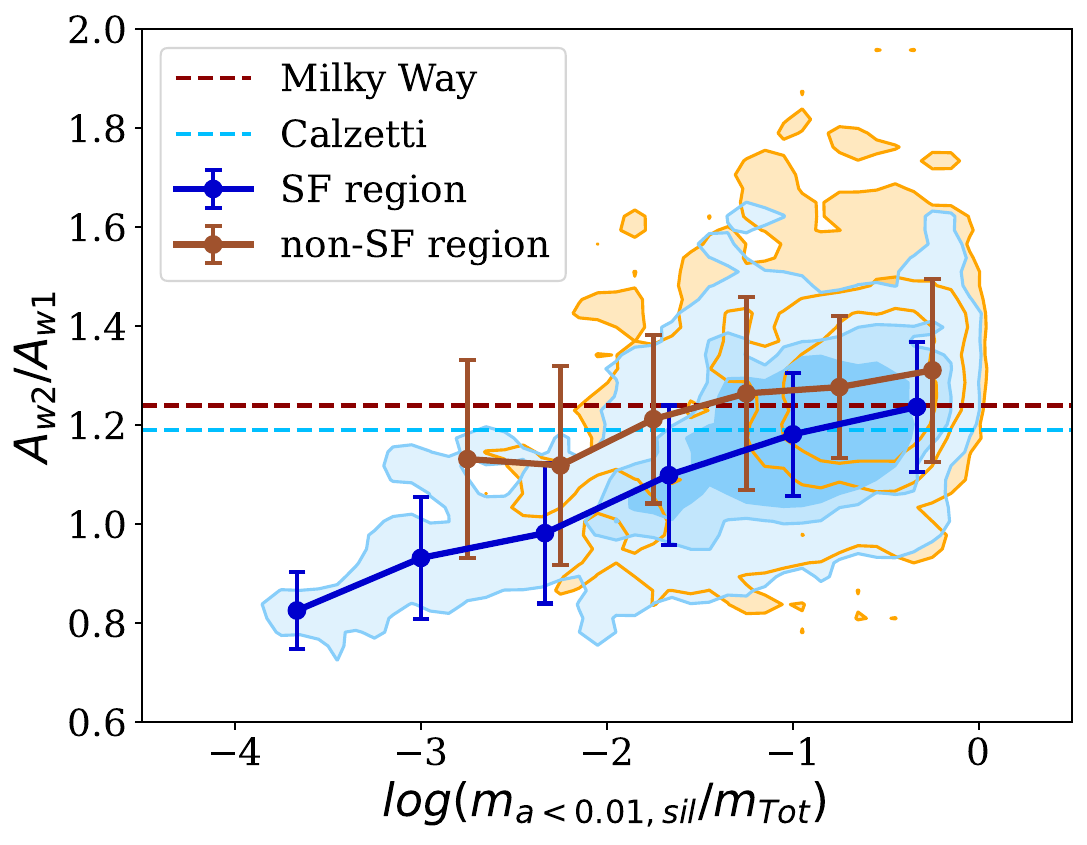}{0.245\textwidth}{}
		\fig{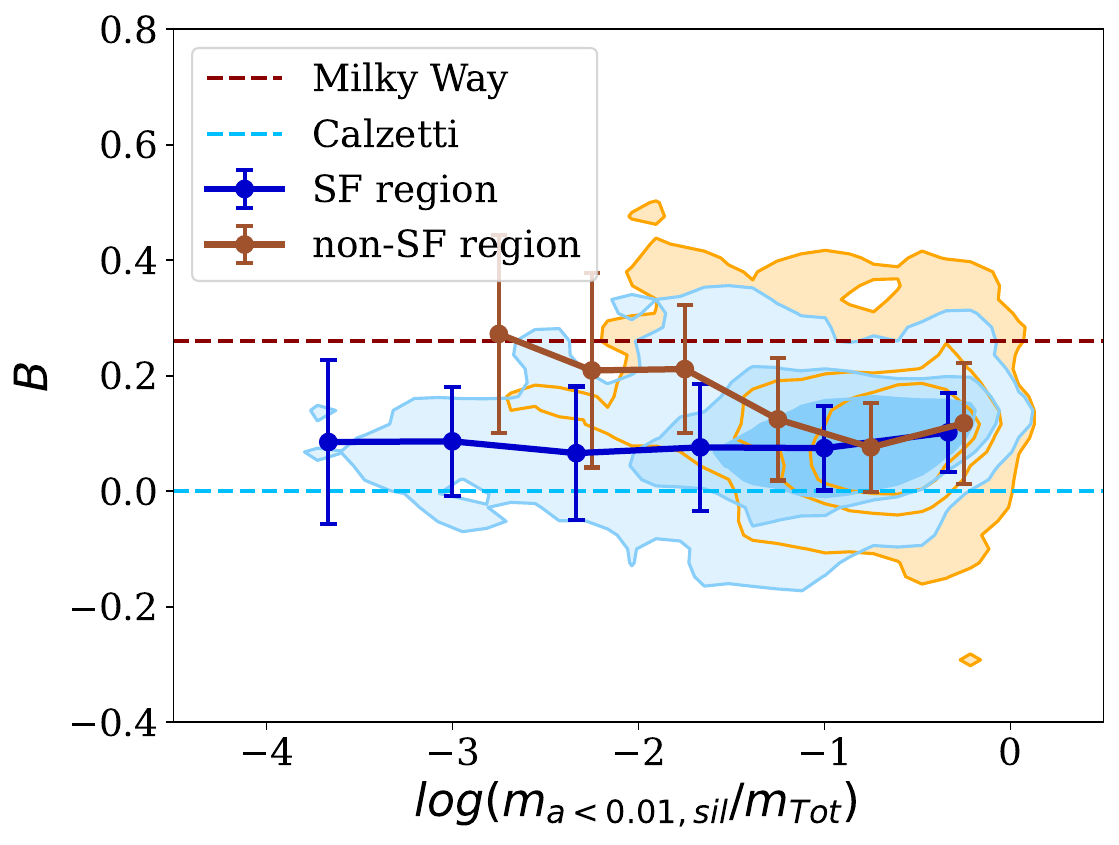}{0.245\textwidth}{}}
    \vspace{-1.6\baselineskip}
	\gridline{\fig{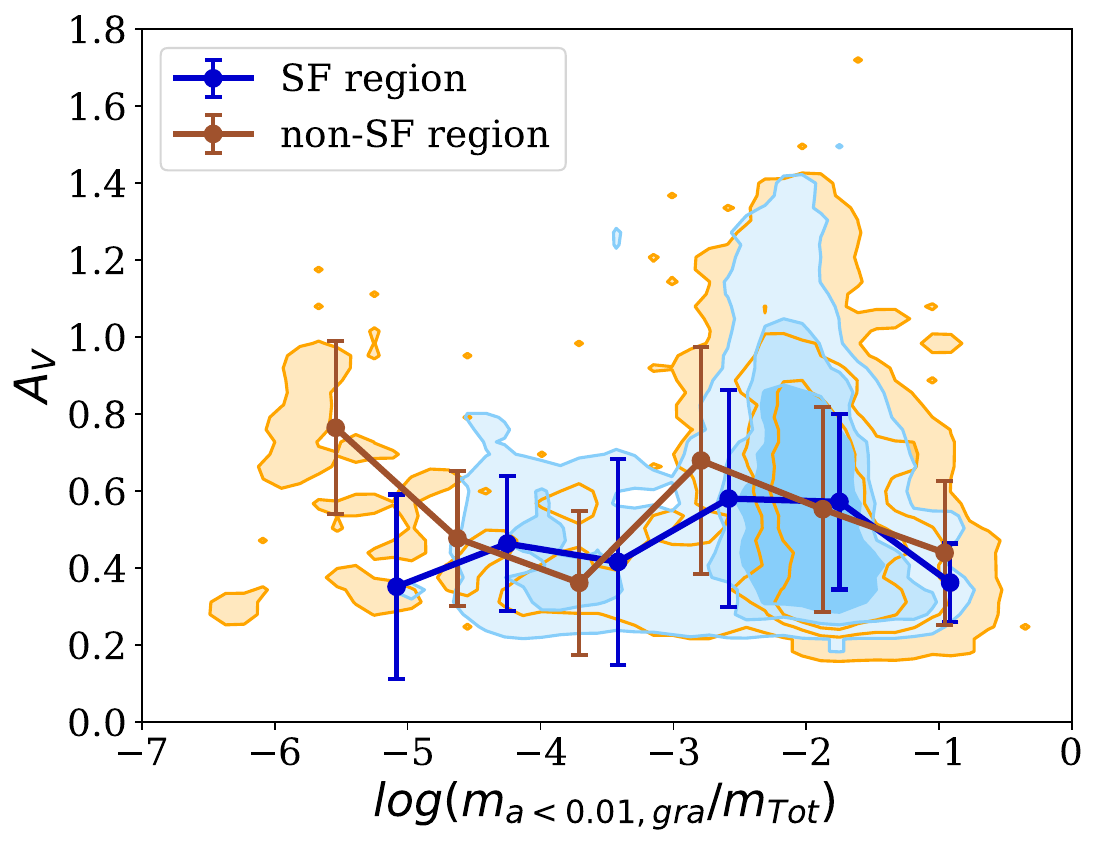}{0.245\textwidth}{}
		\fig{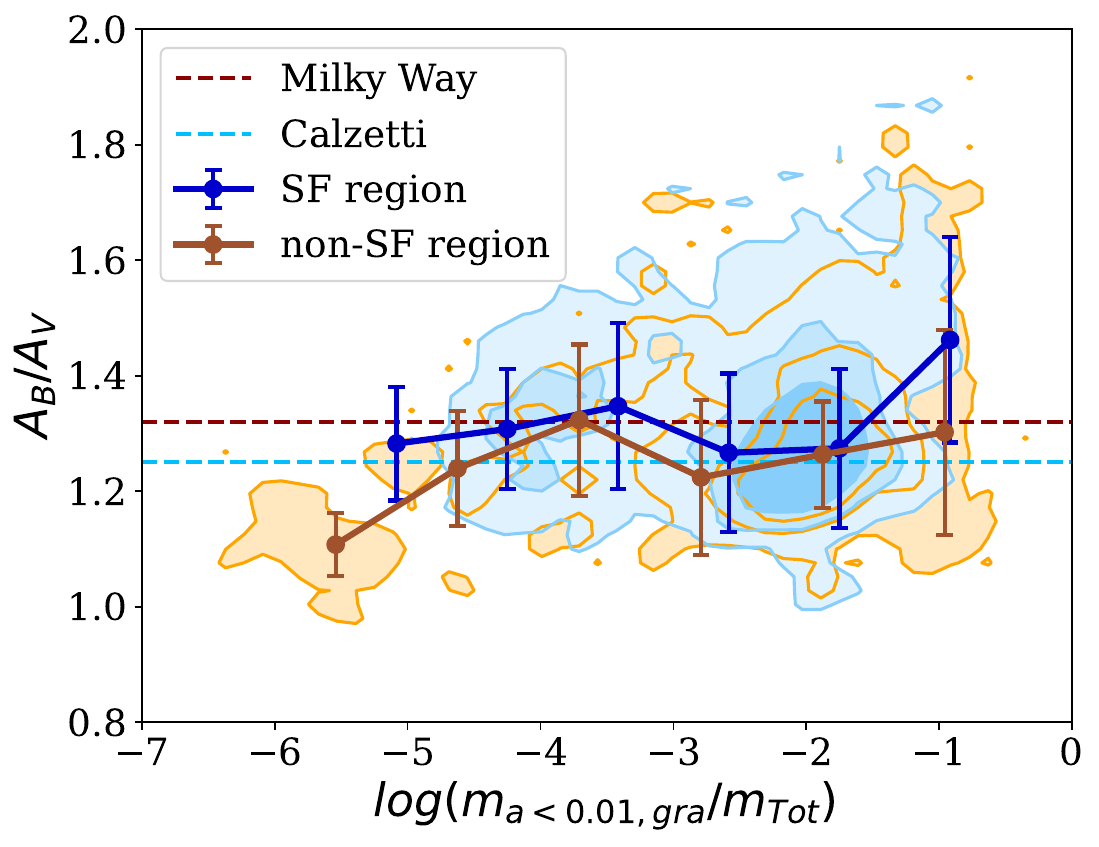}{0.245\textwidth}{}
		\fig{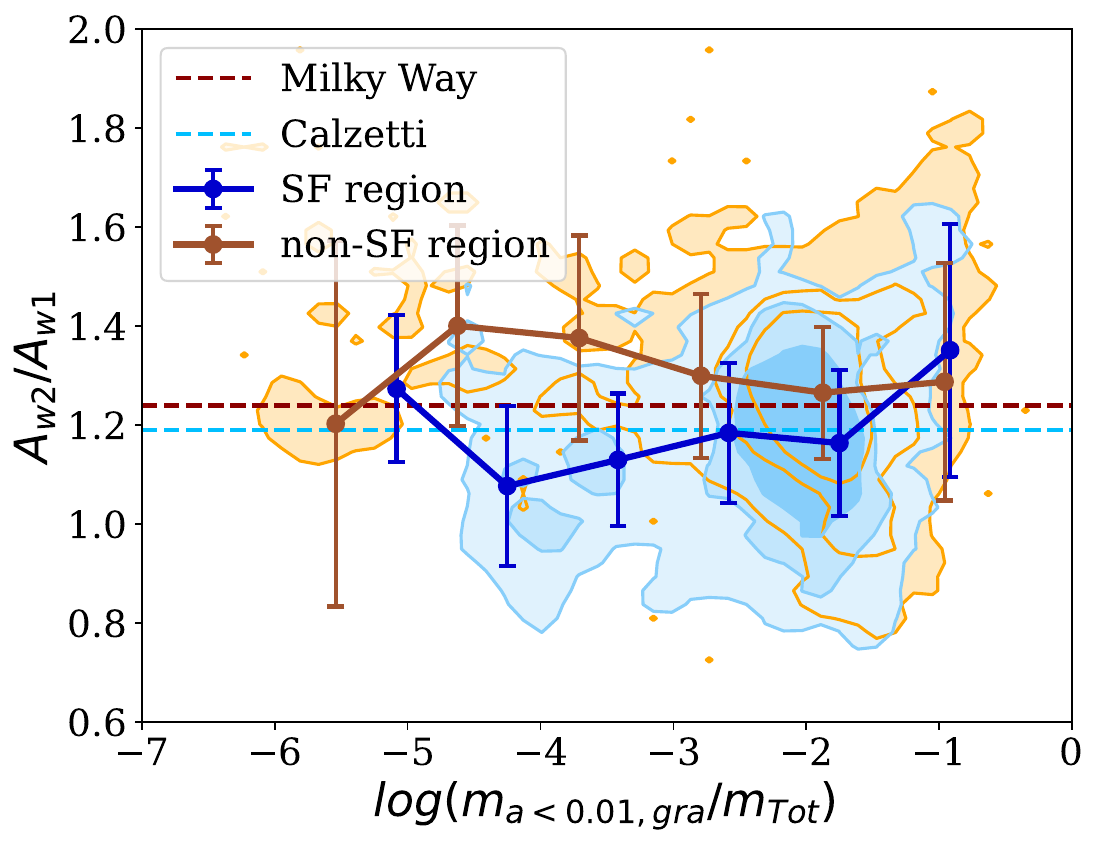}{0.245\textwidth}{}
		\fig{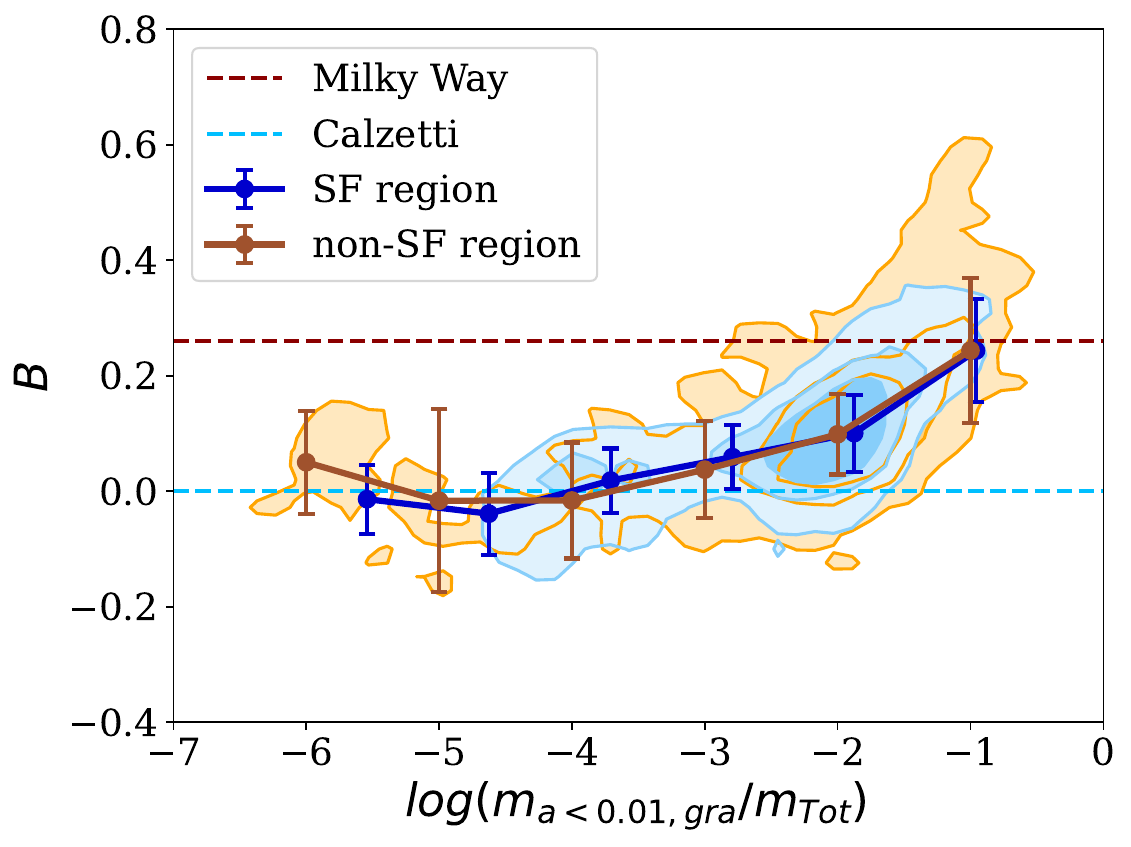}{0.245\textwidth}{}}
	\caption{Relations between observed attenuation properties and model-inferred dust quantities. Columns show, from left to right, $A_V$, $A_B/A_V$, $A_{w2}/A_{w1}$, and the attenuation-curve relative 2175\AA\ bump strength $B$. Rows show, from top to bottom, dust mass surface density, dust-to-stellar mass ratio, silicate mass fraction, small-silicate mass fraction, and small-graphite mass fraction. Light blue contours and blue median curves show SF regions; orange contours and sienna median curves show non-SF regions. Error bars on the median curves are Poisson uncertainties. For $A_B/A_V$, $A_{w2}/A_{w1}$, and $B$, horizontal dotted lines mark reference values from the Milky Way extinction curve \citep{1989ApJ...345..245C} and the Calzetti attenuation curve \citep{2000ApJ...533..682C}.\label{fig:attenuation_dust_related}}
\end{figure*}

\autoref{fig:attenuation_dust_related} compares the observed attenuation properties with the dust quantities inferred from the model. Because the model is fitted directly to the attenuation curves, some relations are expected by construction; their value is to translate the empirical features measured in Papers II and III into model-dependent quantities such as dust column, composition, and small-grain fractions.

The strongest amplitude relation is the positive correlation between $A_V$ and $\Sigma_{\rm dust}$, as expected because $\Sigma_{\rm dust}$ is inferred from the attenuation amplitude after the curve shape is matched. The relation is not strictly proportional, however, because the conversion depends on the fitted dust model. At fixed $A_V$, non-SF regions require larger inferred $\Sigma_{\rm dust}$ than SF regions, possibly reflecting differences in effective dust-star geometry.

The optical slope $A_B/A_V$ shows only weak trends with $\Sigma_{\rm dust}$ and $\Sigma_{\rm dust}/\Sigma_\ast$. The NUV slope $A_{w2}/A_{w1}$ increases more clearly with $\Sigma_{\rm dust}$ in both SF and non-SF regions, but has little dependence on $\Sigma_{\rm dust}/\Sigma_\ast$. The relative bump strength $B$ decreases weakly with $\Sigma_{\rm dust}$, especially in non-SF regions at low dust surface density, and becomes nearly flat at higher $\Sigma_{\rm dust}$. Thus attenuation amplitude alone does not determine the UV curve shape or relative bump strength.

The clearest shape relations involve the small-grain fractions. In both SF and non-SF regions, $B$ increases strongly with $\log_{10}(m_{a<0.01,{\rm gra}}/m_{\rm Tot})$, while $A_{w2}/A_{w1}$ increases with $\log_{10}(m_{a<0.01,{\rm sil}}/m_{\rm Tot})$ and with the total silicate mass fraction. Within this model, the bump is therefore linked most directly to the effective small-graphitic/carbonaceous component, whereas the NUV slope is linked most closely to the small-silicate fraction and overall silicate contribution.

\subsection{\texorpdfstring{Dependence on local H$\alpha$ emission and stellar mass}{Dependence on local H-alpha emission and stellar mass}}

\begin{figure*}[!thbp]
\begin{center}
        \gridline{\fig{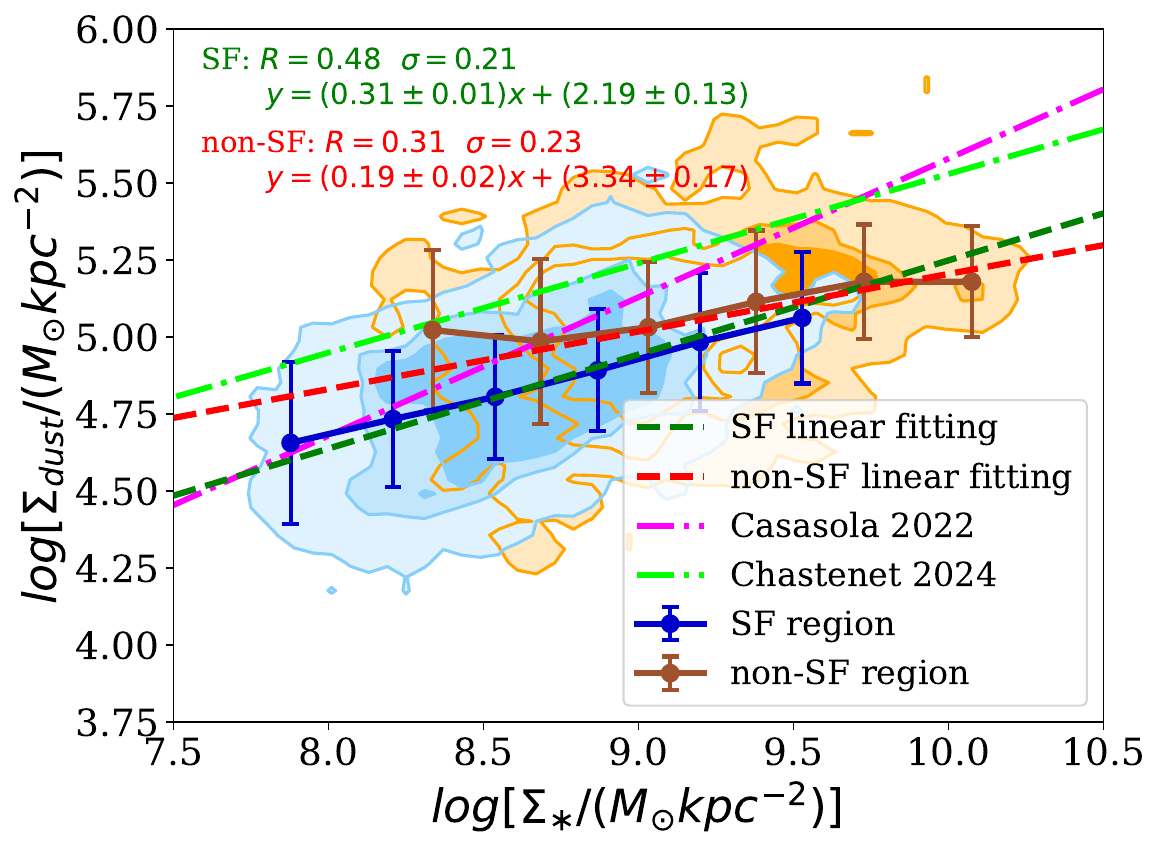}{0.3\textwidth}{}
        \fig{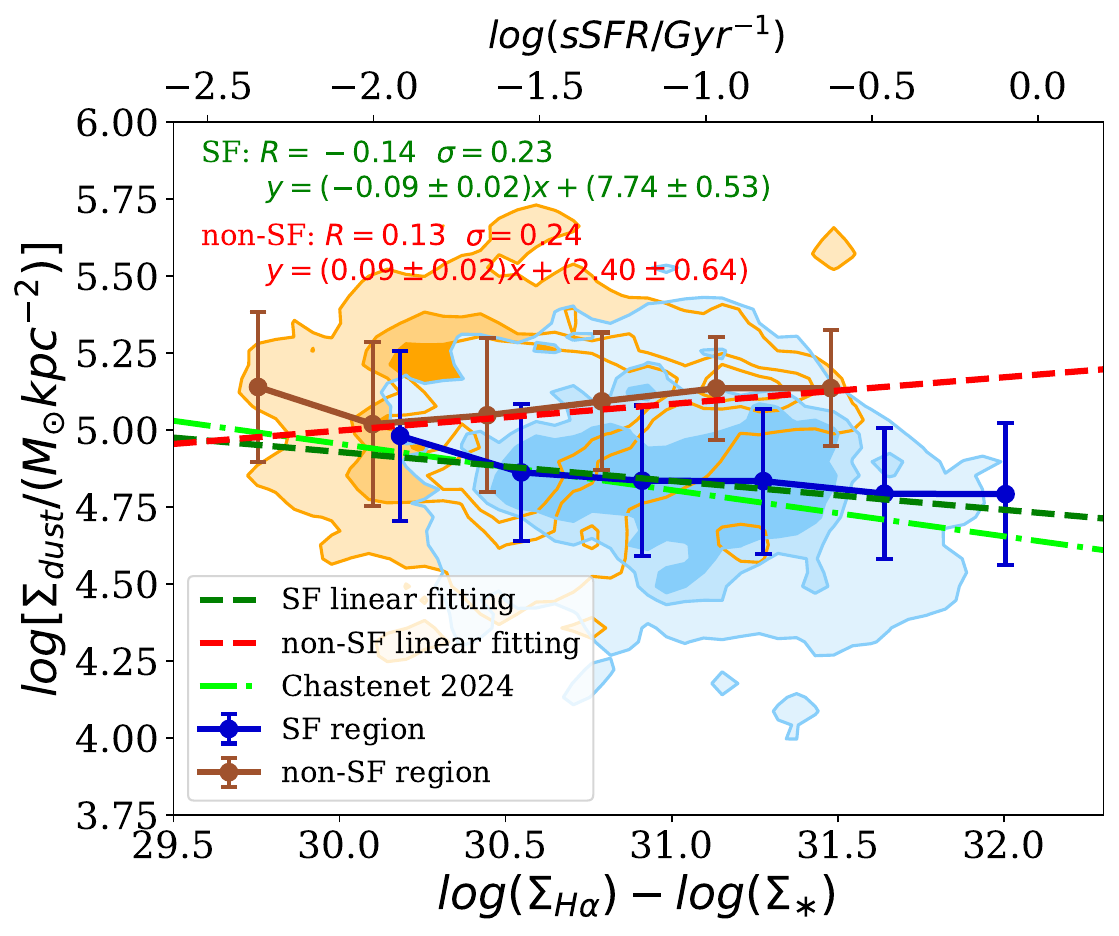}{0.3\textwidth}{}
                \fig{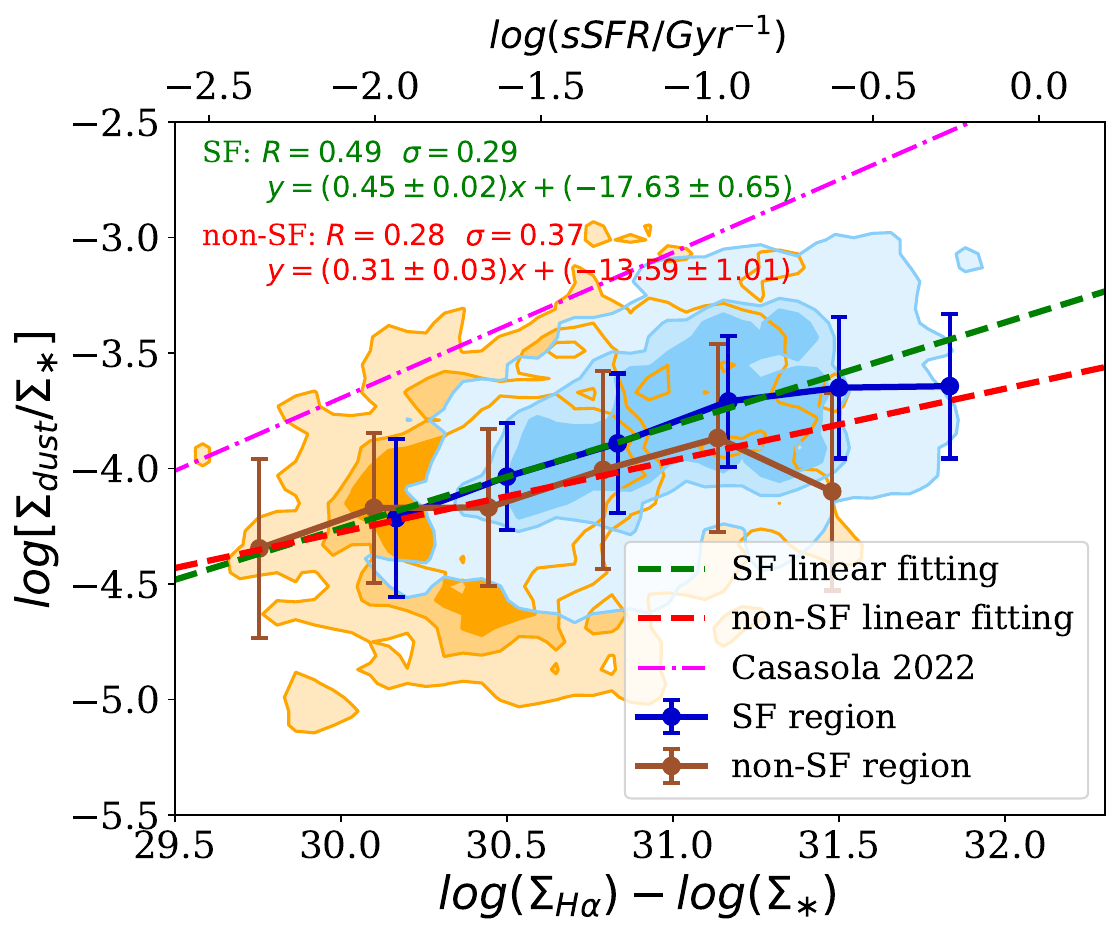}{0.3\textwidth}{}}
    \vspace{-1.8\baselineskip}
        \gridline{\fig{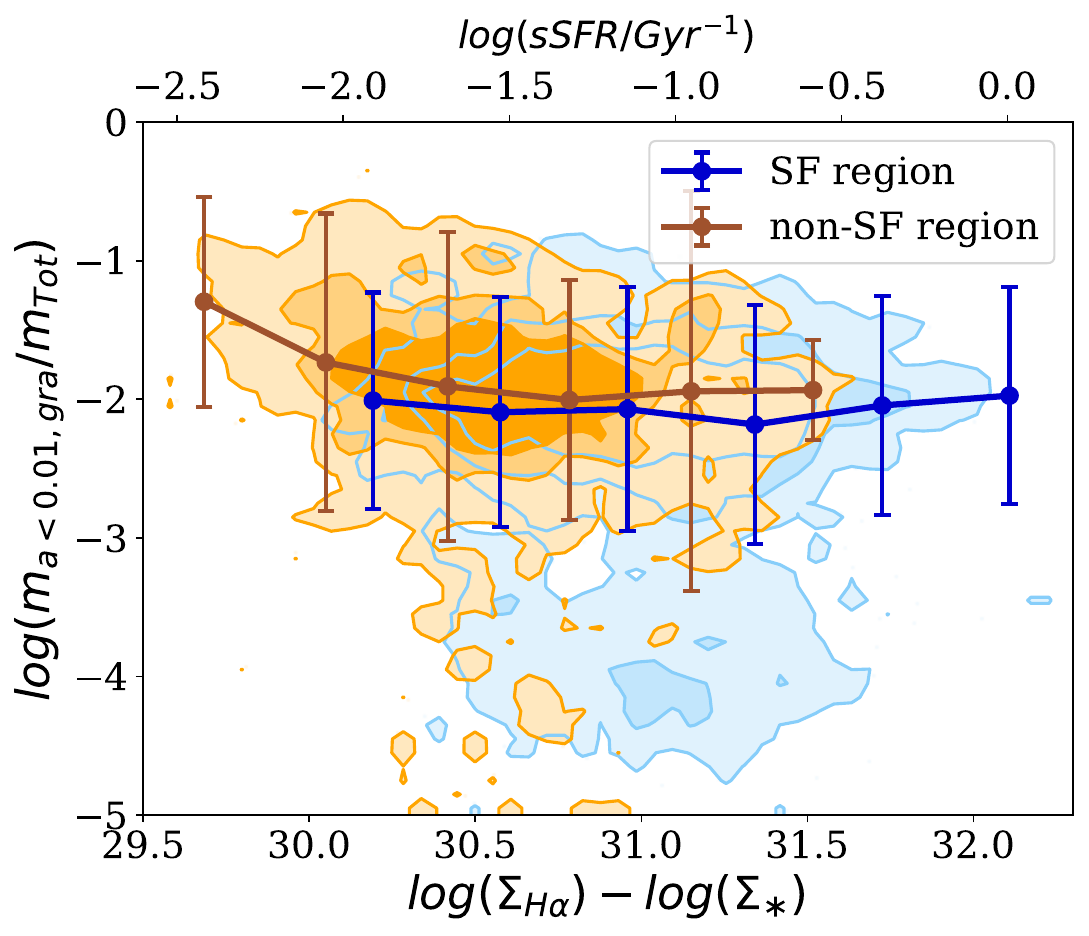}{0.3\textwidth}{}
              \fig{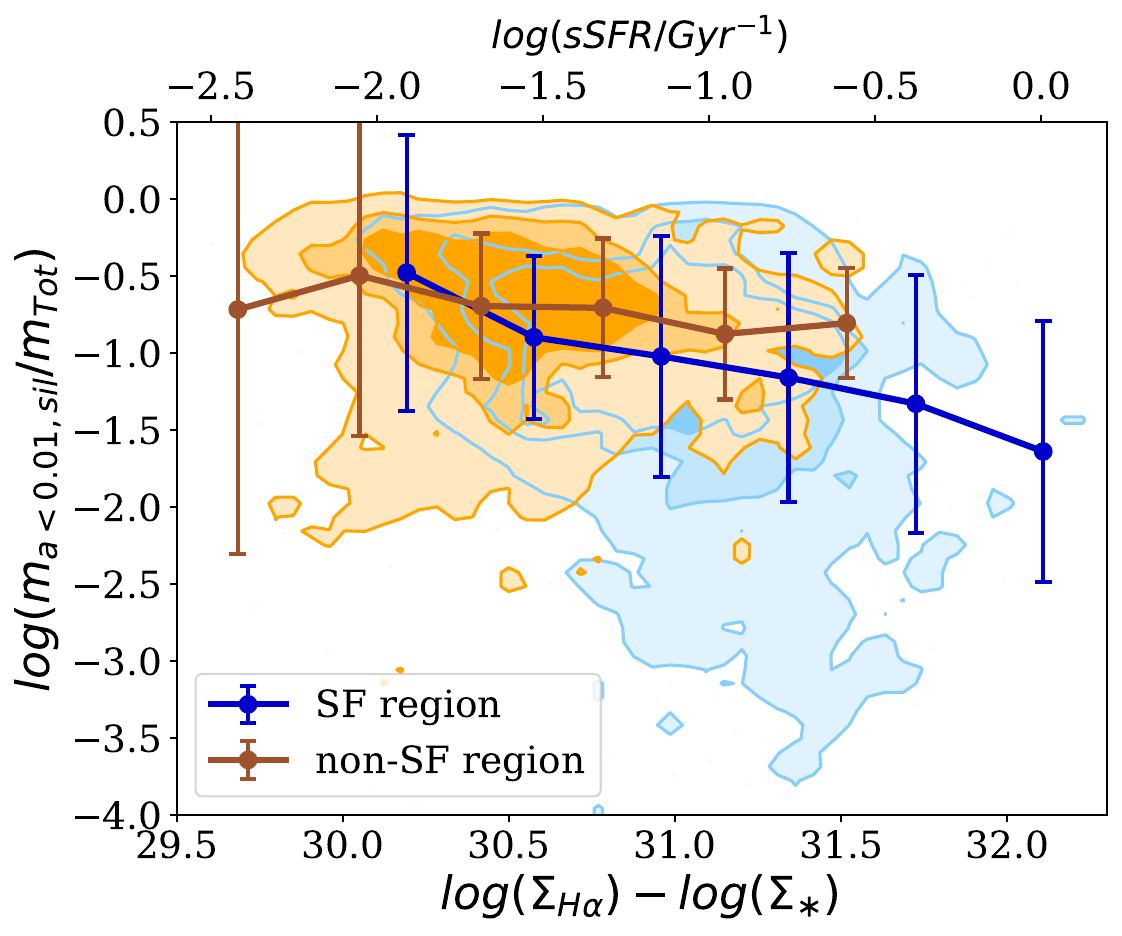}{0.3\textwidth}{}
              \fig{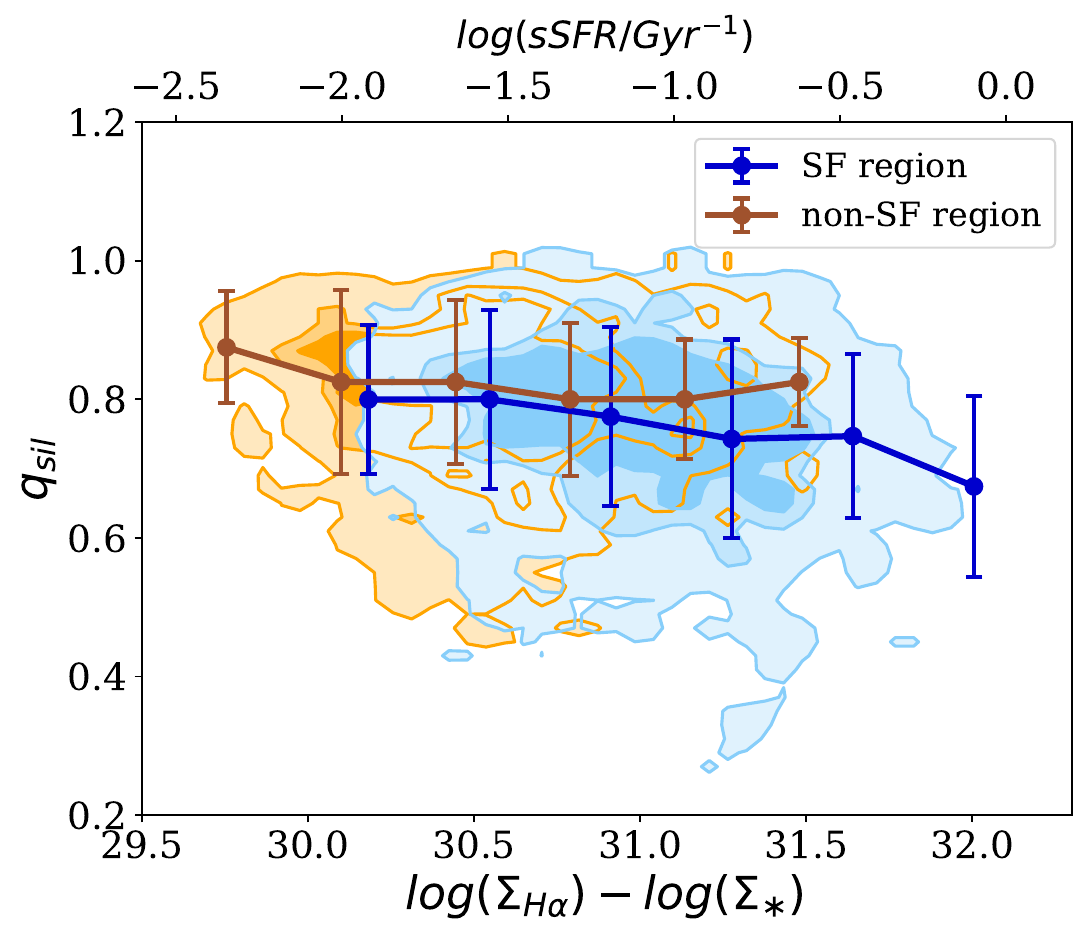}{0.3\textwidth}{}}
    \vspace{-0.5\baselineskip}
        \figcaption{Dust content and grain/composition diagnostics as functions of local stellar mass and H$\alpha$ emission. Top row: dust mass surface density versus stellar mass surface density, dust mass surface density versus specific H$\alpha$ surface brightness $\Sigma_{H\alpha}/\Sigma_\ast$, and dust-to-stellar mass ratio versus $\Sigma_{H\alpha}/\Sigma_\ast$. Bottom row: small-graphite mass fraction, small-silicate mass fraction, and silicate mass fraction versus $\Sigma_{H\alpha}/\Sigma_\ast$. Blue contours and median curves show SF regions; orange contours and sienna median curves show non-SF regions. In the top row, dark green and red dashed lines show linear fits to the SF and non-SF samples, respectively; pink and light-green dot-dashed lines show the resolved DustPedia relations from \citet{2022A&A...668A.130C} and the mid-infrared-based relations from \citet{2025ApJS..276....2C}.\label{fig:DM_DMtoSM_sSFR}}
\end{center}
\end{figure*}

\autoref{fig:DM_DMtoSM_sSFR} compares dust content and grain/composition diagnostics with local stellar mass surface density and specific H$\alpha$ surface brightness, $\Sigma_{H\alpha}/\Sigma_\ast$. In SF regions, $\Sigma_{H\alpha}/\Sigma_\ast$ is closely related to sSFR; in non-SF regions it is interpreted as ionized-gas emission per unit stellar mass. We fit the SF and non-SF samples separately with $y=kx+b$ and refer to correlations with $0\le R<0.3$, $0.3\le R<0.7$, and $R\ge0.7$ as weak, moderate, and strong, respectively.

The top-left panel of \autoref{fig:DM_DMtoSM_sSFR} shows that $\Sigma_{\rm dust}$ increases with $\Sigma_\ast$ in both SF and non-SF regions. Non-SF regions occupy larger stellar and dust surface densities overall. The fitted slopes are $k=0.31$ for SF regions and $k=0.19$ for non-SF regions, with correlation coefficients of $R=0.48$ and $R=0.31$, respectively. Thus local stellar surface density is a significant, but not exclusive, predictor of the attenuation-inferred dust column.

The top-middle panel shows only weak trends between $\Sigma_{\rm dust}$ and $\Sigma_{H\alpha}/\Sigma_\ast$. The fitted slope is slightly negative in SF regions ($k=-0.09$) and slightly positive in non-SF regions ($k=0.09$), but both correlation coefficients are below 0.3. Thus the absolute dust mass surface density is not primarily controlled by $\Sigma_{H\alpha}/\Sigma_\ast$ in this sample.

By contrast, the dust-to-stellar mass ratio in the top-right panel increases with $\Sigma_{H\alpha}/\Sigma_\ast$. The trend is moderate in SF regions ($R=0.49$) and weaker in non-SF regions ($R=0.28$). This indicates that the dust content per unit stellar mass is more closely connected to local recent star formation, or to local ionized-gas activity in non-SF regions, than the absolute dust column is.

The external relations overplotted in the top row provide context rather than direct calibrations for our attenuation-based dust masses. The resolved DustPedia relations of \citet{2022A&A...668A.130C} and the mid-infrared-based relations of \citet{2025ApJS..276....2C} use emission-based dust measurements and different sample selections, so order-of-magnitude agreement is more meaningful than exact agreement in slope or normalization.

The bottom row examines how the inferred small-grain and composition parameters vary with $\Sigma_{H\alpha}/\Sigma_\ast$. The small-graphite fraction is low and nearly flat across SF regions, apart from a subset of especially low values; in non-SF regions it decreases more clearly with increasing $\Sigma_{H\alpha}/\Sigma_\ast$. The small-silicate fraction decreases with $\Sigma_{H\alpha}/\Sigma_\ast$ in both ionization classes, with a steeper decline in SF regions, and the silicate mass fraction follows a similar trend. The same dust quantities have weaker or less coherent dependence on $\Sigma_{H\alpha}$ and $\Sigma_\ast$ separately, suggesting that the trends reflect the balance between ionized-gas activity and the underlying stellar mass distribution.

\subsection{Dependence on recent star-formation history}

\begin{figure*}[!thbp]
\gridline{\fig{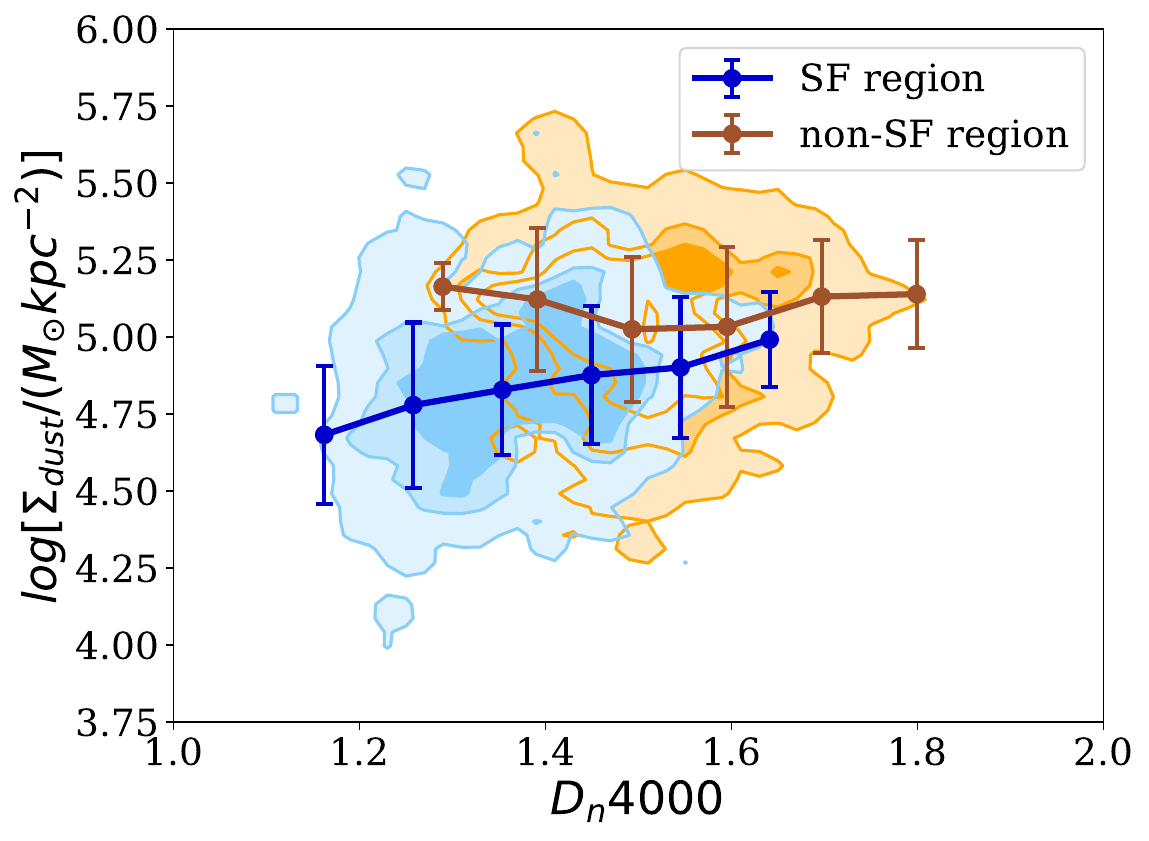}{0.3\textwidth}{}
\fig{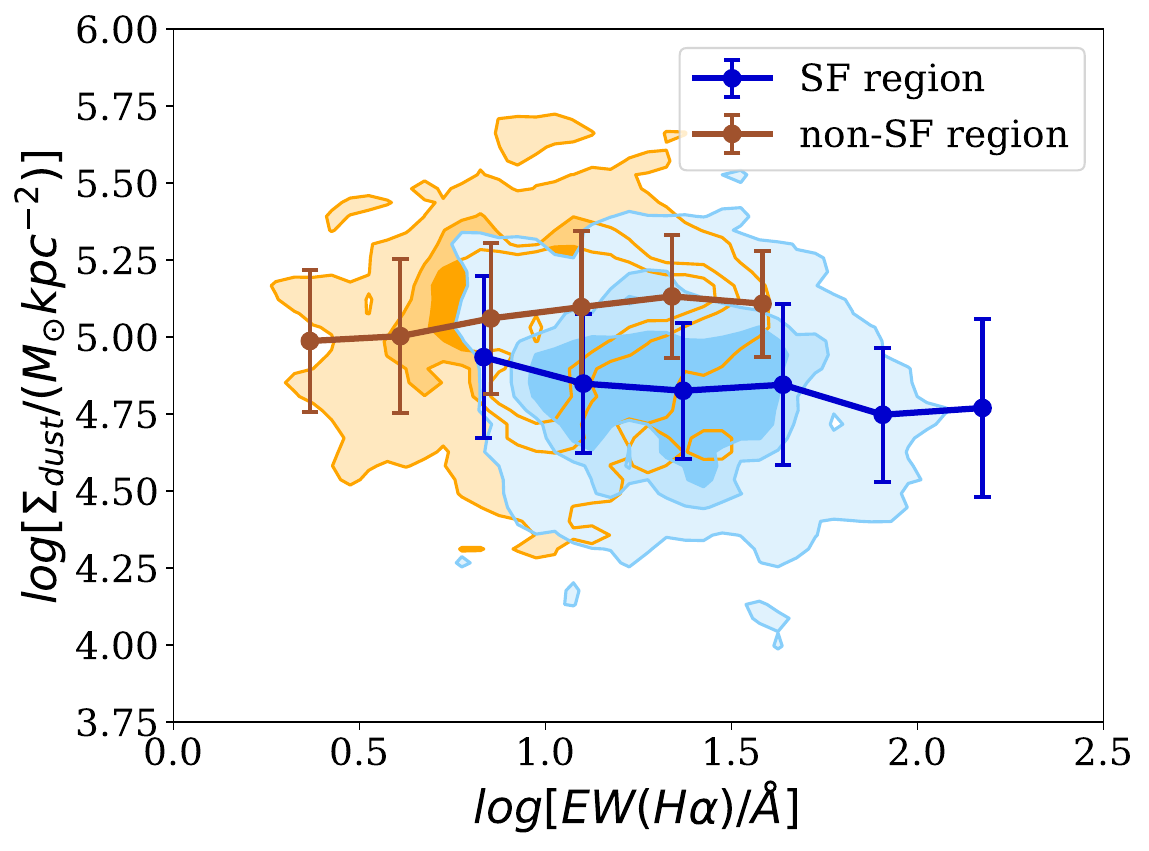}{0.3\textwidth}{}
\fig{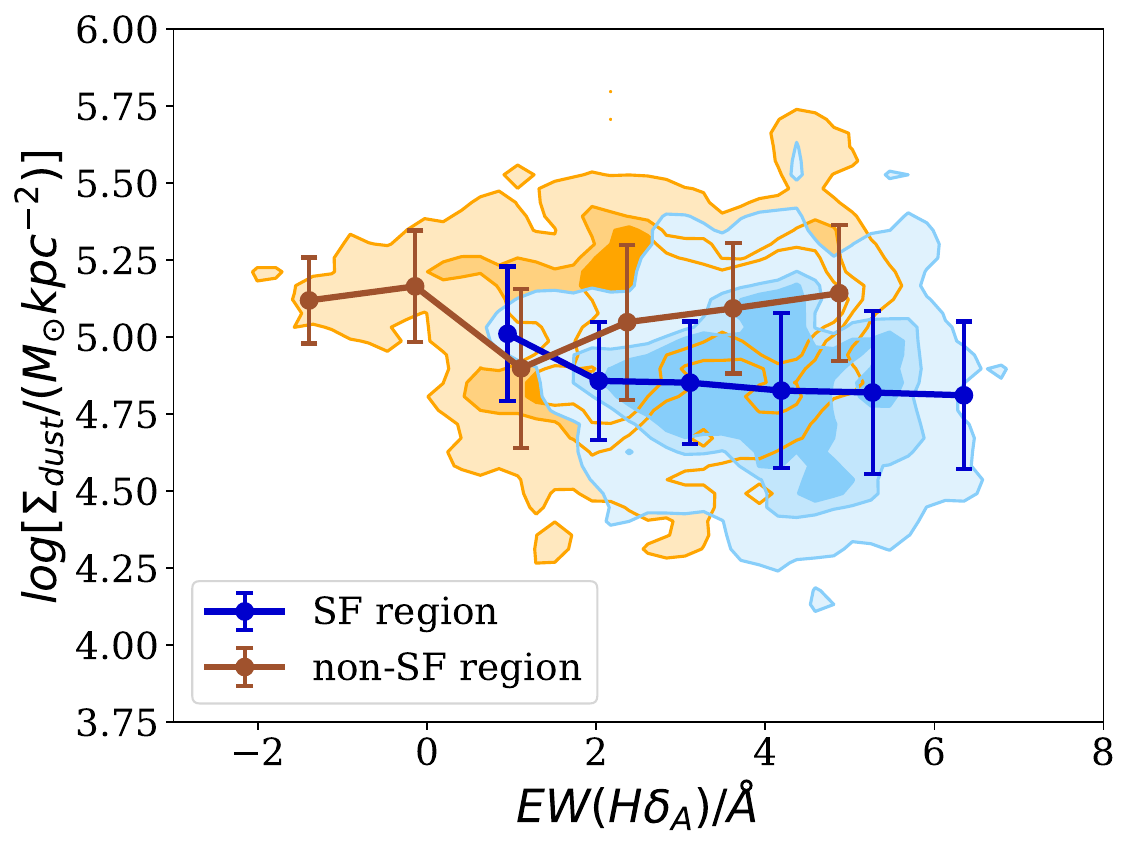}{0.3\textwidth}{}}
\vspace{-1.8\baselineskip}
	\gridline{\fig{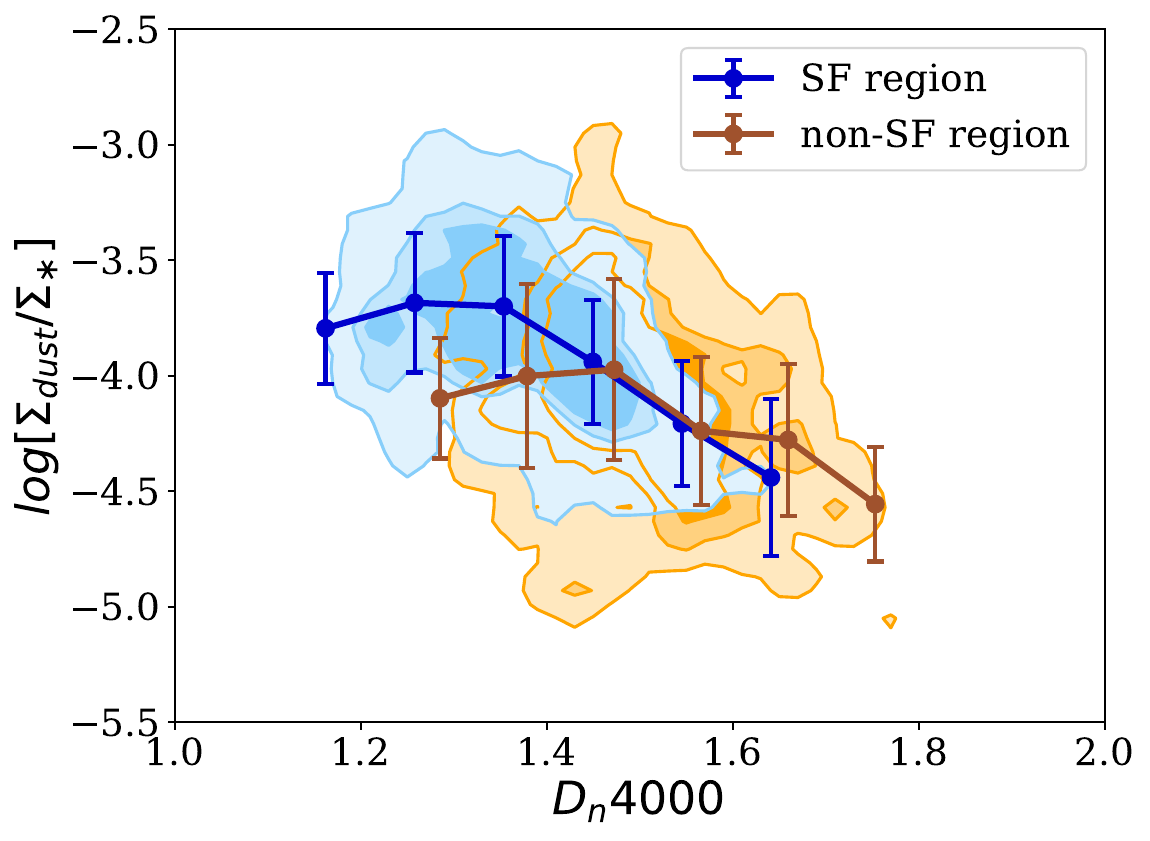}{0.3\textwidth}{}
		      \fig{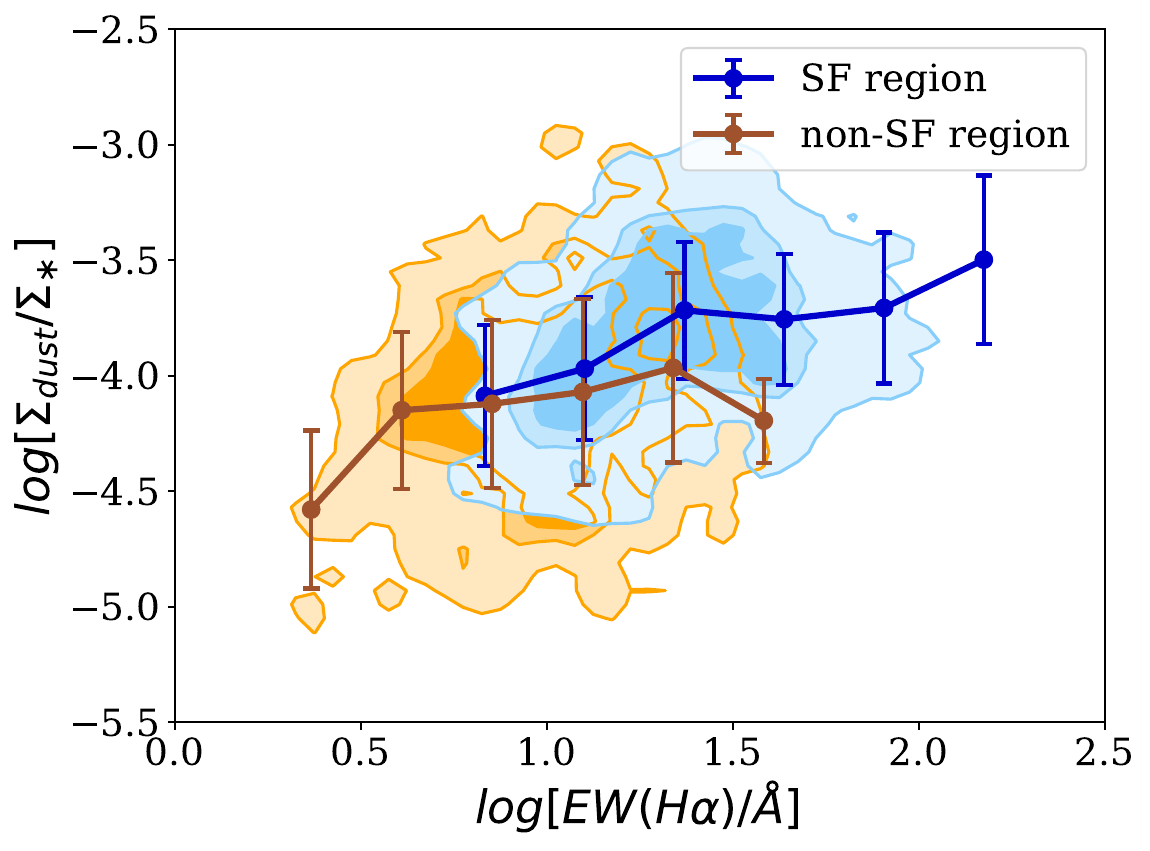}{0.3\textwidth}{}
\fig{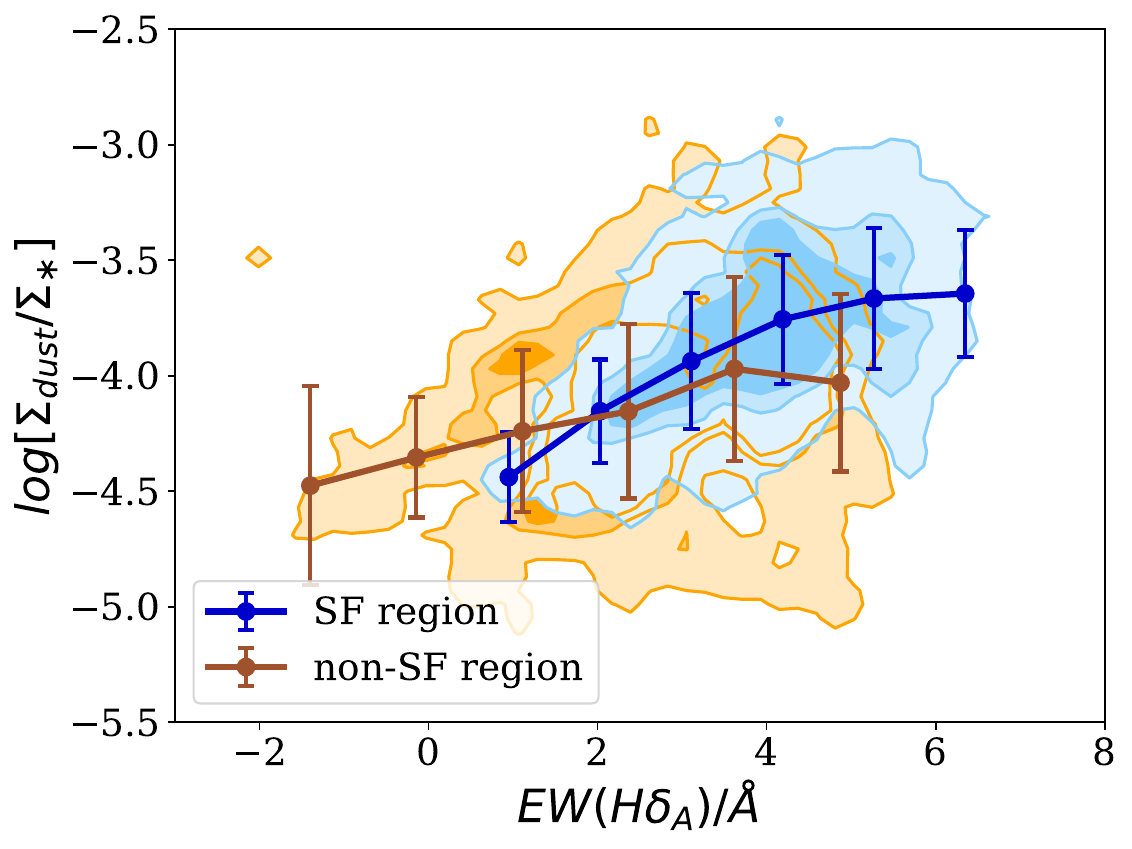}{0.3\textwidth}{}}
	\caption{Dust content as a function of recent-SFH diagnostics. The upper row shows dust mass surface density $\Sigma_{\rm dust}$, and the lower row shows dust-to-stellar mass ratio $\Sigma_{\rm dust}/\Sigma_\ast$. Columns, from left to right, show $D_n4000$, EW(H$\alpha$), and EW(H$\delta_{\rm A}$). Symbols, contours, median curves, and colors are the same as in \autoref{fig:DM_DMtoSM_sSFR}.\label{fig:DM_DMtoSM_SFH}}
\end{figure*}

\autoref{fig:DM_DMtoSM_SFH} compares dust content with three recent-SFH diagnostics: $D_n4000$, EW(H$\alpha$), and EW(H$\delta_{\rm A}$). Larger $D_n4000$ indicates older stellar populations, EW(H$\alpha$) traces very recent massive-star formation in SF regions, and EW(H$\delta_{\rm A}$) is sensitive to star formation over the past several hundred Myr. In the upper row, $\Sigma_{\rm dust}$ increases with $D_n4000$ and decreases with EW(H$\alpha$) and EW(H$\delta_{\rm A}$) in SF regions; in non-SF regions, it shows little dependence on these diagnostics.

The lower row shows the opposite behavior for $\Sigma_{\rm dust}/\Sigma_\ast$: in both SF and non-SF regions, the dust-to-stellar mass ratio decreases with $D_n4000$ and increases with EW(H$\alpha$) and EW(H$\delta_{\rm A}$). Together with the bottom row of \autoref{fig:DM_DMtoSM_sSFR}, these trends emphasize a contrast between total dust content and grain-size diagnostics: regions with stronger recent activity have more dust per unit stellar mass, but lower relative small-grain abundance.

\subsection{\texorpdfstring{Secondary dependences on stellar age and metallicity}{Secondary dependences on stellar age and metallicity}}

We also examined stellar-age and metallicity indicators as secondary checks. The age trends are consistent with the recent-SFH diagnostics: $\Sigma_{\rm dust}/\Sigma_\ast$ decreases with stellar age, while the inferred small-grain fractions are higher in older regions. Metallicity mainly affects the absolute dust column: $\Sigma_{\rm dust}$ increases with stellar metallicity and, in SF regions, with gas-phase oxygen abundance. By contrast, the inferred small-grain fractions and silicate mass fraction show no strong metallicity dependence over the metal-rich range covered by this sample. Because these trends largely repeat the SFH sequence or provide consistency checks rather than defining a separate branch of the argument, we summarize them here without retaining additional figure grids.

\section{Discussion}\label{sec:discussion}

\subsection{What the attenuation-based dust model adds} \label{subsec:discuss_method}

Dust masses in galaxies are most often inferred from infrared to submillimeter emission, either through modified-blackbody estimates or through full SED fitting with physical dust models \citep[e.g.][]{1990A&A...237..215D, 2007ApJ...657..810D, 2011A&A...525A.103C, 2017A&A...602A..46J, 2020A&A...636A..18R, 2023ApJ...948...55H}. These methods directly use dust thermal emission and can constrain dust temperature and luminosity, but their inferred masses depend on dust temperature, opacity, cold-dust treatment, and wavelength coverage \citep[e.g.][]{1983QJRAS..24..267H, 2020RSOS....700556H, 2021MNRAS.508L..58B, 2023MNRAS.520.5506C}. Far-infrared facilities such as Herschel have enabled resolved dust studies in nearby galaxies, while ALMA provides high-resolution continuum measurements for more distant systems, often with more limited temperature constraints.

Our approach is complementary. We do not fit dust thermal emission; instead, we use the UV-to-NIR attenuation curves measured in Papers I and II and interpreted empirically across Papers I--III. By fitting the normalized curve shape first, we allow the effective grain-size distribution and silicate-to-graphite mixture to vary from spaxel to spaxel, and then use the attenuation amplitude to estimate a model-dependent dust mass surface density. The strength of the method is therefore its controlled translation of the observed attenuation-curve sequence into a physically interpretable dust-model framework. Its limitation is equally important: attenuation curves are effective quantities that also depend on dust-star geometry, scattering, and stellar-population mixing. The absolute values of $\Sigma_{\rm dust}$ should therefore be interpreted as model-dependent estimates, whereas the more robust information lies in relative trends and in the connection between fitted grain parameters and observed attenuation-curve features.

\subsection{Dust content on kiloparsec scales} \label{subsec:discuss_previous_dust_mass}

Large galaxy samples show that dust mass, stellar mass, star formation rate, and metallicity follow broad scaling relations, and that the dust-to-stellar mass ratio is closely linked to stellar mass and sSFR \citep[e.g.][]{2012A&A...540A..52C, 2017MNRAS.464.4680D, 2019A&A...623A...5D, 2020A&A...633A.100C, 2020MNRAS.496.3668D}. Our resolved results are qualitatively consistent with this picture. Non-SF regions have larger absolute dust mass surface densities, reflecting their larger stellar mass surface densities, but SF regions have larger dust-to-stellar mass ratios. The increase of $\Sigma_{\rm dust}/\Sigma_\ast$ with $\Sigma_{H\alpha}/\Sigma_\ast$ and its decrease with stellar age are the local, kiloparsec-scale counterparts of the global connection between specific dust content and recent star formation.

Comparisons with resolved emission-based studies should remain qualitative. High-resolution SED fitting and DustPedia studies show that dust, gas, metallicity, and star-formation quantities remain correlated on sub-galactic scales, but that the detailed slopes depend on the adopted modeling assumptions and physical scale \citep[e.g.][]{2011A&A...536A..88G, 2022A&A...668A.130C}. We therefore do not expect our attenuation-based dust masses to match emission-based calibrations one-to-one. Agreement with external relations should be viewed as an order-of-magnitude sanity check rather than as an absolute calibration of the dust-mass scale. The metallicity trends should be interpreted with the same caution: $\Sigma_{\rm dust}$ increases with stellar and gas-phase metallicity in SF regions, as expected if metal-rich regions have larger dust reservoirs, but $\Sigma_{\rm dust}/\Sigma_\ast$ is more strongly affected by the underlying dependence on $\Sigma_\ast$. Because most of our spaxels are metal-rich, the weak dependence of the fitted small-grain fractions on metallicity does not rule out stronger dust-evolution effects at lower metallicity.

\subsection{Small grains, UV slopes, and the 2175\AA\ bump} \label{subsec:discuss_previous_size_distribution}

The fitted grain-size parameters provide the most direct link between this paper and Papers I--III. Paper II showed that the NUV attenuation slope varies strongly with local stellar-population and emission-line properties, and Paper III showed that the relative 2175\AA\ bump strength is strongest in regions with weak recent-star-formation or ionized-gas activity. Paper III also separated absolute bump amplitude from normalized bump prominence, showing that the former partly follows dust column whereas the latter better traces the effective prominence of the bump carriers. Within the adopted model, these empirical trends translate into model parameters: the small-graphite fraction increases with the attenuation-curve relative bump strength $B$, while the small-silicate fraction and total silicate mass fraction increase with the NUV slope $A_{w2}/A_{w1}$. These correlations are partly built into the model physics, but they show that the observed attenuation-curve variations can be represented by coherent changes in effective small-grain fractions.

This interpretation is broadly consistent with dust-evolution calculations in which stellar sources, shattering, accretion, coagulation, and destruction reshape the grain-size distribution over time \citep[e.g.][]{2013MNRAS.432..637A, 2017MNRAS.466..105A, 2019MNRAS.485.1727H, 2020MNRAS.491.3844A}. Our sample does not probe the low-metallicity regime where models predict the strongest changes in dust-to-metal and small-to-large grain ratios, but the high-metallicity SF spaxels with very low small-grain fractions are qualitatively consistent with efficient coagulation or destruction in dense, active environments. Observationally, the narrow global range and sSFR scatter of small-to-large grain ratios \citep{2022MNRAS.515.5306R}, and the decline of PAH fraction with sSFR or star-formation surface density \citep{2025ApJS..276....2C}, are also consistent with small carbonaceous or aromatic grains being processed in regions with stronger recent star formation.

\subsection{A local dust-processing picture} \label{subsec:discuss_physics}

Taken together, the results support a coherent extension of Papers I--III. Papers I--III showed that attenuation-curve variations on kiloparsec scales are governed mainly by local conditions rather than by global inclination or radius. Here we find that the same local sequence can be expressed in dust-model terms: regions with larger $\Sigma_{H\alpha}/\Sigma_\ast$ tend to have larger dust-to-stellar mass ratios but smaller inferred small-grain fractions. In SF regions, recent massive-star formation can enrich the ISM through stellar winds, supernova ejecta, and subsequent grain growth, while associated radiation fields, shocks, and dense-gas processing can reduce the relative abundance of small grains. This combination naturally produces more dust per unit stellar mass while weakening the small-grain signatures that shape the NUV slope and the 2175\AA\ bump. Recent THESAN-ZOOM simulations likewise find that bursty star formation and stellar feedback can destroy or eject dust from star-forming regions and suppress dust survival \citep{2026arXiv260708824G}, although they focus on high-redshift galaxies rather than nearby kiloparsec-scale regions.

The interpretation of non-SF regions requires extra care. As emphasized in Paper III, $\Sigma_{H\alpha}/\Sigma_\ast$ is not a direct sSFR in non-SF spaxels; it is an empirical measure of ionized-gas emission per unit stellar mass and may include contributions from diffuse ionized gas, evolved stars, shocks, weak AGN activity, or mixtures of these sources. The decline of small-grain fractions with this quantity suggests that grain processing is not confined to classical HII regions, but the physical driver may differ from that in SF regions. We therefore interpret the non-SF trends as empirical links between dust properties and ionized-gas activity, not as direct evidence for ongoing star-formation-driven processing alone. The weaker ongoing massive-star dust production in non-SF regions may also help explain why they have lower dust-to-stellar mass ratios overall.

The graphite and silicate components do not vary identically. In SF regions the small-graphite fraction is already low and changes only weakly with $\Sigma_{H\alpha}/\Sigma_\ast$, while in non-SF regions it declines more clearly from the lowest ionized-gas-emission regime, mirroring the Paper III result that the relative 2175\AA\ bump is strongest in the most quiescent regions. The small-silicate fraction decreases more continuously with $\Sigma_{H\alpha}/\Sigma_\ast$, especially in SF regions, suggesting that the NUV attenuation slope is especially sensitive to this component. This interpretation remains model-dependent: the uniform-screen geometry, two-component silicate+graphite composition, and MRN-like size distributions cannot uniquely determine carrier chemistry or true dust geometry. Nevertheless, the model achieves the main goal of this paper by translating the attenuation-curve trends established in Papers I--III into a physically interpretable picture in which local small-grain processing links optical/NUV attenuation slopes, the 2175\AA\ bump, and resolved dust content.

\section{Summary}\label{sec:summary}

This is the fourth paper in a series studying dust attenuation at kiloparsec scales in nearby galaxies. We interpret the attenuation curves measured in Paper II, using the method developed in Paper I, with a simple silicate+graphite dust model for 2487 high-continuum-S/N spaxels in 91 SwiM v4.2 galaxies. The model assumes MRN-like grain-size distributions and a uniform-screen geometry. The resulting quantities are effective, model-dependent dust properties, but they provide a physical translation of the attenuation-curve trends established in the series. Our main conclusions are as follows.

\begin{enumerate}
  \item A two-component silicate+graphite model can reproduce the main observed attenuation-curve shapes. The best-matched models are generally silicate dominated, and non-SF regions have a higher inferred silicate mass fraction than SF regions.
  \item Non-SF regions have larger dust mass surface densities than SF regions, but lower dust-to-stellar mass ratios. This difference separates absolute dust content from dust content per unit stellar mass: non-SF regions are denser in stars and dust, whereas SF regions contain more dust relative to their local stellar mass.
  \item The model parameters provide a physical bridge to Papers I--III. The inferred small-graphite fraction correlates strongly with the attenuation-curve relative 2175\AA\ bump strength $B$, while the small-silicate fraction and total silicate mass fraction correlate with the NUV slope $A_{w2}/A_{w1}$. Thus the attenuation-curve variations measured in Papers II and III can be interpreted, within this model, as changes in the effective relative abundance of small grains.
  \item The dust-to-stellar mass ratio increases with specific H$\alpha$ surface brightness, $\Sigma_{H\alpha}/\Sigma_\ast$, in both SF and non-SF regions. As in Paper III, this quantity is interpreted as sSFR-like in SF regions and as ionized-gas emission per unit stellar mass in non-SF regions. It also decreases with luminosity- and mass-weighted stellar age. The absolute dust mass surface density shows weaker trends with $\Sigma_{H\alpha}/\Sigma_\ast$ but increases with stellar metallicity and, in SF regions, with gas-phase metallicity. Thus specific dust content is the more sensitive tracer of local recent star-formation or ionized-gas activity, while metallicity mainly enters through the absolute dust column.
  \item The small-grain fractions generally decrease with increasing $\Sigma_{H\alpha}/\Sigma_\ast$, but the behavior differs between components and ionization classes. In SF regions the small-graphite fraction is low and changes weakly, while the small-silicate fraction decreases clearly. In non-SF regions both small-graphite and small-silicate fractions decline with $\Sigma_{H\alpha}/\Sigma_\ast$. These trends are consistent with local processing or destruction of small grains in regions with stronger radiation fields or ionized-gas activity.
  \item Overall, the model-dependent results support the series picture that resolved attenuation-curve variations are primarily shaped by local conditions. The 2175\AA\ bump, the NUV slope, and the inferred small-grain fractions all point to local dust processing on kiloparsec or smaller scales, while metallicity and dust column mainly modulate the strength of these trends.
\end{enumerate}

Future work should test this effective attenuation-based picture against independent dust tracers. Joint modeling of resolved attenuation curves, infrared dust emission, and more flexible radiative-transfer geometries will be needed to separate grain-size variations from dust-star geometry and to calibrate the absolute dust masses. Larger samples and sub-kiloparsec data will show whether the local small-grain processing inferred here remains the dominant driver of attenuation-curve diversity below the MaNGA resolution scale.

\section*{Data Availability}

The MaNGA data products used in this work are publicly available as part of SDSS DR17. The Swift/UVOT+MaNGA value-added catalog is available from the SDSS value-added catalog archive, and the 2MASS imaging data are available from the NASA/IPAC Infrared Science Archive. The derived dust-model catalog, selection masks, and scripts needed to reproduce the figures are available from the corresponding authors on reasonable request.

\begin{acknowledgments}
This work is supported by the National Science Foundation of China (grant No. 12433003) and the National Key R\&D Program of China (grant No. 2018YFA0404502).

Funding for SDSS-IV has been provided by the Alfred P. Sloan Foundation and Participating Institutions. Additional funding for SDSS-IV has been provided by the US Department of Energy Office of Science. SDSS-IV acknowledges support and resources from the Center for High-Performance Computing at the University of Utah. The SDSS web site is www.sdss.org.

We acknowledge the Tsinghua Astrophysics High-Performance Computing platform at Tsinghua University for providing computational and data storage resources that have contributed to the research results reported within this paper.
\end{acknowledgments}

\begin{contribution}


RG performed the data analysis, produced the figures, and wrote the first draft of the manuscript. CL initiated the project, supervised the analysis, and edited the manuscript. TJ, SZ, NL, and ZC contributed to the interpretation of the results and reviewed the manuscript.


\end{contribution}

%
\facilities{Sloan(MaNGA), Swift(UVOT), 2MASS}

\software{Astropy \citep{2013A&A...558A..33A,2018AJ....156..123A,2022ApJ...935..167A}, 
          BHMIE \citep{1983bhwd.book.....B}, 
          BIGS \citep{2019MNRAS.485.5256Z}}

\bibliography{bibtex}{}
\bibliographystyle{aasjournalv7}



\end{document}